\documentclass[aps,prl,reprint,superscriptaddress]{revtex4-2}
\usepackage{epsfig}
\usepackage{graphicx}
\usepackage[caption=false]{subfig}
\usepackage{bm}
\usepackage{color}
\usepackage{float}
\usepackage{mathtools}
\usepackage{amsmath}
\usepackage{comment}

\begin{document}

\newcommand{\NaFeCuAs}{NaFe$_{1-x}$Cu$_{x}$As}
%\preprint{}


\title{RKKY coupled local moment magnetism in NaFe$_{1-x}$Cu$_{x}$As}

%\title{Coexistence of Localized and Itinerant Magnetism in NaFe$_{1-x}$Cu$_{x}$As Single Crystals ($x\approx0.01$)}


%\thanks{}
\author{Yizhou Xin}
\altaffiliation{Corresponding author. Email: yizhouxin2018@\\u.northwestern.edu}
\affiliation{Department of Physics and Astronomy, Northwestern University, Evanston IL 60208, USA}
\author{Ingrid Stolt}
\affiliation{Department of Physics and Astronomy, Northwestern University, Evanston IL 60208, USA}
\author{Yu Song}
\altaffiliation{Present address: Center for Correlated Matter and Department of Physics, Zhejiang University, Hangzhou, China}
\affiliation{Department of Physics and Astronomy and Rice Center for Quantum Materials, Rice University, Houston TX 77005, USA}

\author{Pengcheng Dai}
\affiliation{Department of Physics and Astronomy and Rice Center for Quantum Materials, Rice University, Houston TX 77005, USA}
\author{W. P. Halperin}
\affiliation{Department of Physics and Astronomy, Northwestern University, Evanston IL 60208, USA}


\date{\today}
\begin{abstract}
A central question in a large class of strongly correlated electron systems, including heavy fermion compounds and iron pnictides, is the identification of different phases and their origins. It has been shown that the antiferromagnetic (AFM) phase in some heavy fermion compounds is induced by Ruderman-Kittel-Kasuya-Yosida (RKKY) interaction between localized moments, and that the competition between this interaction and Kondo effect is responsible for quantum criticality. However,  conclusive experimental evidence of the RKKY interaction in pnictides is lacking. Here, using high resolution $^{23}$Na NMR measurements on lightly Cu-doped metallic single crystals of \NaFeCuAs ~($x \approx 0.01$) and numerical simulation, we show direct evidence of the RKKY interaction in this pnictide system. Aided by computer simulation, we identify the $^{23}$Na NMR satellite resonances with the RKKY oscillations of spin polarization at Fe sites. Our $^{23}$Na spin-lattice and spin-spin relaxation data exhibits signature of an itinerant and inhomogeneous AFM phase in this system, accompanied by a simulation of Cu-induced perturbation to the ordered moments on the Fe sites. Our NMR results indicate coexistence of local and itinerant magnetism in lightly Cu-doped \NaFeCuAs.

\end{abstract}

%\keywords{}

\maketitle
\section{Introduction}

% we show that antiferromagnetism arises from

In the early days of investigation of the Kondo effect, Boyce and Slichter~\cite{Boy.74} and Alloul~\cite{All.74,All.77}, studied magnetic interactions of a dilute magnetic impurity, such as Fe,  in a non-magnetic, metallic Cu host.  They observed $^{63}$Cu NMR satellites which they identified with atomic positions of Cu atoms in shells at varying distances from the impurity, attributed to the Ruderman-Kittel-Kasuya-Yosida (RKKY) interaction of local moments at the Fe sites, coupling through the hyperfine field to the copper nuclei. This interaction is mediated by the conduction electron spins. The satellites determine both the range and the magnitude of the RKKY oscillations in the electronic spin density. It has been shown by past studies that this interaction plays an important role in the formation of magnetic order in strongly correlated electron systems. For instance, in some heavy fermion metals, the antiferromagnetic (AFM) order is driven by the RKKY interaction among the localized $f$-electrons, and is proposed to compete with the Kondo effect to be responsible for quantum criticality~\cite{Pas.21,Var.76,Geg.08}. For iron chalcogenides, \textit{ab initio} electronic structure calculations show that this interaction stabilizes bicollinear AFM order~\cite{Hir.15}. 

Numerical studies of iron pnictides based on multiband models~\cite{Akb.11} show that the RKKY interaction becomes anisotropic in EuFe$_2$As$_2$ where suppression of superconductivity has been reported~\cite{Zap.11}. Magnetic suppression of superconductivity has also been attributed to RKKY-like interactions in the vicinity of Mn impurities in LaFeAsO$_{1-x}$F$_{x}$ and  BaFe$_{2}$As$_{2}$ compounds~\cite{Mor.17, Leb.14}. In these instances, $^{75}$As NMR spectral shifts are associated with the $^{75}$As nuclei that are nearest-neighbor (NN) to the Mn impurities. However, as far as we know, there is no direct experimental evidence for existence of the RKKY interaction in pnictides.  It is important to explore if RKKY can be directly observed in iron pnictides. To this end, we take advantage of high resolution $^{23}$Na NMR, which we use to identify the manifestation of the RKKY interaction. This is what we report here.

\begin{figure}
	\includegraphics[scale=0.45]{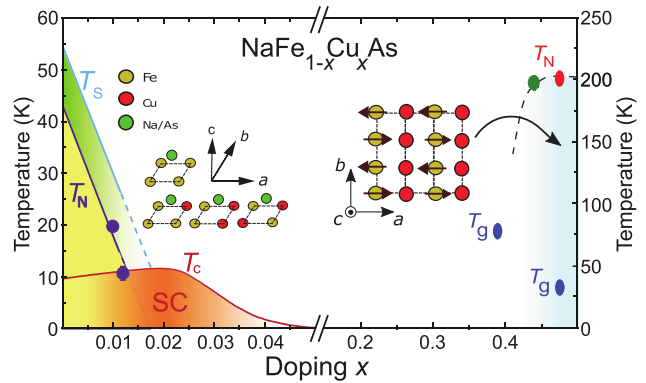}
	\caption{\label{PhaseDiagram} Magnetic order in NaFe$_{1-x}$Cu$_{x}$As single crystals; phase diagram adapted from Refs ~\cite{Son.15,Xin.20}. The system evolves from the lightly Cu-doped regime, to a correlated insulator with heavy  doping. For $x > 0.3$, the green data point is from neutron scattering~\cite{Son.16}; red and blue data points are from nuclear magnetic resonance (NMR) measurements~\cite{Xin.20}. $T_\textrm{N}$ is the N\'{e}el temperature for antiferromagnetic transitions, and $T_\textrm{g}$ is the temperature of spin-glass transitions~\cite{Xin.20}. The two data points in the lightly Cu-doped regime are from this work. $T_\textrm{s}$ and $T_{c}$ are the temperatures of tetragonal-to-orthorhombic structural and superconducting transitions, respectively. Insets show Na and As sites relative to nearest-neighbor (NN) Fe or Cu substituted for Fe.}
\end{figure}

%%introduction to previous work on NaFeCuAs
The Cu-doped pnictide, NaFe$_{1-x}$Cu$_{x}$As, is distinct from other iron pnictides. It is a Mott insulator in the heavily doped regime ($x\approx 0.5$), while for low copper doping concentrations ($x \leq 0.05$), it is metallic, exhibiting superconductivity in close proximity to an AFM ordered state~\cite{Wan.13,Ma.11}. At high doping long-range AFM order arises from localized moments on the Fe sites~\cite{Son.16, Mat.16,Zha.17,Xin.19, Xin.20} and the Cu dopants are nonmagnetic in a valence state Cu$^{1+}$ with completely filled 3$d$ orbitals~\cite{Son.16,Son.21}. We have previously studied this regime~\cite{Xin.19, Xin.20} using  $^{23}$Na NMR showing that long range order is only achieved for $x$ very close to 0.5; otherwise, the magnetic state is a spin glass as exemplified in the case $x=0.39$ [Fig.~\ref{PhaseDiagram}]. In contrast, for low concentrations $x$ as we report here, the Cu dopants are found to be electron doped, suggesting that they are magnetic with valence 2+~\cite{Wan.13,Cui.13}. The magnetic Fe ions are in the same valence state as the dopants~\cite{Son.21}. In this case, it is not yet clear whether antiferromagnetism follows a local moment picture or is itinerant, associated with Fermi surface nesting~\cite{Yi.12,Zha.10,Maz.08,Nin.10,Joh.10,Si.08,Yil.08,Kou.09}.

%%Try to answer this question, what makes this compounds ideal for seaching for RKKY? 
The magnetic Cu impurities in a metallic host makes lightly Cu-doped NaFe$_{1-x}$Cu$_{x}$As an ideal system for searching for the effects of the RKKY interaction. Two NMR-sensitive nuclei, $^{23}$Na and $^{75}$As, are located on opposite sides of the Fe layer (Fig.~\ref{PhaseDiagram}). Both nuclei are coupled to spin polarization of the Fe ions through transferred hyperfine interaction, providing complementary information about the magnetism in the conducting plane. However, one distinct advantage of choosing the $^{23}$Na nucleus for NMR measurements lies in that it is associated with a significantly narrower NMR linewidth than $^{75}$As due to its weaker hyperfine coupling~\cite{Xin.19}, allowing higher spectral resolution in probing distributions of local magnetic fields. 

%% what is our discovery? 
Using $^{23}$Na NMR measurements on lightly Cu-doped metallic single crystals of NaFe$_{1-x}$Cu$_{x}$As ($x\approx 0.01$), we have discovered direct evidence for the RKKY interaction in a pnictide system for the first time. Our $^{23}$Na NMR spectra of the central transition (1/2 $\leftrightarrow$ -1/2) reveal four satellites, which can be similarly interpreted as was the case for dilute alloys of Fe in Cu. However, there are significant differences with the Kondo system analogy. For our pnictide crystals, the pure host material exhibits antiferromagnetism with a small ordered magnetic moment, $m \approx 0.17~\mu_B/\textrm{Fe}$, and transition temperature $T_\textrm{AF} \approx 46$\,K~\cite{Li.09}. Assuming a random distribution of Cu-dopants, we have simulated the corresponding hyperfine field distribution at the Na sites at room-temperature and compared with our experiment, identifying each of the $^{23}$Na NMR satellites with a shell of Na nuclei. This leads to a value for the Fermi wave vector, $k_{F}$, consistent with other measurements~\cite{Akb.13}, and accounts for the frequencies and spectral weight of the satellite resonances.

Measurements of spin-lattice ($1/T_{1}$) and spin-spin ($1/T_{2}$) relaxation rates provide characterization of the electronic spin dynamics which, in the former case, is at nonzero wave-vector. Measuring $1/T_{1}T$ at low temperatures close to the AFM transition $T_{\mathrm{AF}}$, we find NMR signatures indicative of 3D spin fluctuations in a weak itinerant AFM metal~\cite{Mor.74}. These results in lightly Cu-doped NaFe$_{1-x}$Cu$_{x}$As are complemented by simulation, and are consistent with our interpretation of an itinerant  inhomogeneous AFM phase coexisting with local magnetism, very different from the more heavily doped, non-metallic, crystals of the same compound $0.5 >x > 0.3$~\cite{Xin.20}.

%%%%%%%%%%%%%%%%%  Phase Diagram  %%%%%%%%%%%%%%%%%%
%%%%%%%%%%%%%%%%%%%%%%%%%%%%%%%%%%%%%%%%%%%%%%%%%%%%%

%%%%%%%%%%%%%%%%%   F I G U R E  1   %%%%%%%%%%%%%%%%%%
\begin{figure*}
	\hspace*{-0.5cm}
	\includegraphics[scale = 0.15]{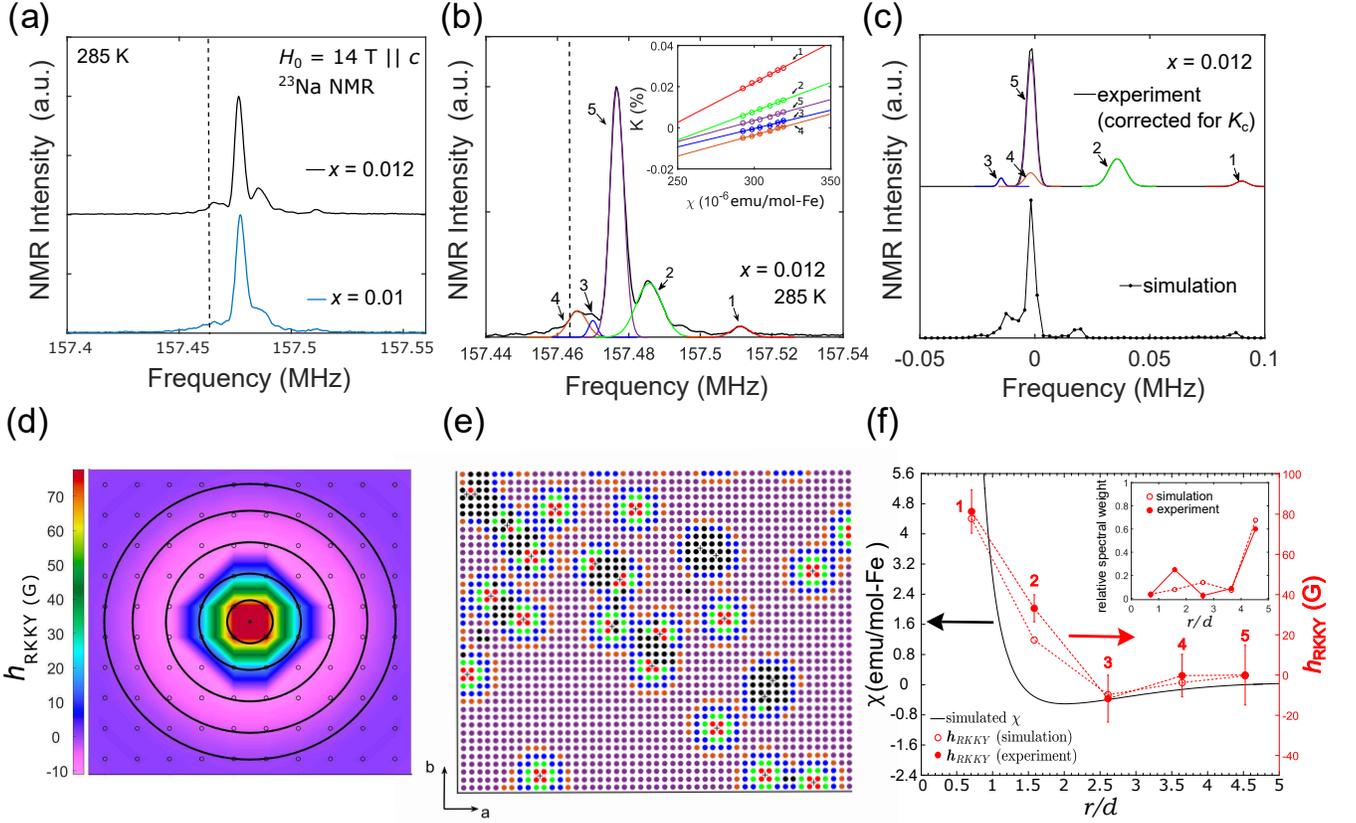}
	\caption{\label{Fig.1} RKKY interactions for $x \approx 0.01$ (a) Room-temperature $^{23}$Na spectra of NaFe$_{1-x}$Cu$_{x}$As for the central transition (-1/2 $\leftrightarrow$ 1/2), $H_{0}$ = 14 T $||~c$-axis for $x = 0.010$ and 0.012. (b) For $x = 0.012$, Gaussian fits of spectral components indicated by color. The inset, $K(T)$, versus bulk susceptibility, $\chi(T)$, $125 ~\textrm{K} \leq T \leq 250 ~\textrm{K}$ for each spectral component; the slopes are the hyperfine field form factors. The numbers correspond to different spectral components. (c) Comparison of the RKKY model with the Gaussian fits to the experimental spectrum, corrected for the chemical shift, $K_{c}$, obtained from b).  (d) Calculated hyperfine fields in the $ab$-plane, $h_{\textrm{RKKY}}$. Na sites are small black open circles near the Cu dopant (black cross).  Large black circles label shells of Na sites  color coded in b), c) and e). (e) Larger view of the simulation field. (f) Black trace is the susceptibility $\chi_{\textrm{RKKY}}$ versus the ratio of the radial distance $r$ from a Cu dopant to the lattice spacing $d$ between two NN Fe atoms. The  data points are measurements of hyperfine fields for Na in different shells. The numbers correspond to the different spectral components in b) and c). The inset shows a comparison of relative spectral weights between the simulated and experimental results for each spectral component.}
\end{figure*}
%%%%%%%%%%%%%%%%%%%%%%%%%%%%%%%%%%%%%%%%%%%%%

\section{Experimental Methods}
The single crystals of NaFe$_{1-x}$Cu$_{x}$As which we performed NMR measurements on were grown by the self-flux method at Rice University~\cite{Son.16}. Detailed information on sample preparation, including specially designed hermetic sample holders for NMR measurements, can be found elsewhere~\cite{Xin.19}. We performed $^{23}$Na NMR experiments for compounds $x$ = 0.010 and 0.012, with an external magnetic field, $H_0 = 13.98$ T, along the sample $c$-axis, the [001] direction. Since $^{23}$Na has a nuclear spin of 3/2, we used the quadrupolar echo pulse sequence ($\pi/2$-$\tau$-$\pi/2$-$\tau$-echo) to maximize the excitation bandwidth for collecting NMR spectrum. The linewidth is taken to be the full width at half maximum (FWHM) of a spectrum. For $T_{2}$ measurements, we varied the delay time, $\tau$, from 0.2 $\mu$s to 2 ms to measure recovery profile from which $T_{2}$ was extracted by fitting.  A saturation-recovery sequence was used for $T_{1}$ measurements. The length of $\pi/2$ pulse length was typically set to $\sim6\,\mu s$.  
%%%%%%%%%%%%%%%%%%%%%%%%%%%%%%%%%%%%%%%%%%%%%%%%%%%%%%%%%%%

\section{RKKY interaction shown in $^{23}$N\lowercase{a} Nuclear Magnetic Resonance Spectra}
The room-temperature $^{23}$Na spectra of the central transition (1/2~$\leftrightarrow$~-1/2) are given in Fig.~\ref{Fig.1}\,(a) for $x = 0.010$ and 0.012.   The quadrupolar satellites ($\pm$3/2~$\leftrightarrow$~$\pm$1/2) are well-resolved and an order of magnitude outside the spectral range shown here~\cite{Xin.19}. The dashed line represents the Larmor frequency, $f_{L}$, calculated from $f_{L}=\gamma H_{0}/2\pi$, where $\gamma$ is the gyromagnetic ratio of $^{23}$Na and $\gamma/2\pi = 11.2625 \mathrm{~MHz/T}$. Both spectra show a larger spectral component with some smaller satellites. The dopant concentrations are sufficiently small that statistically they can be treated as independent impurities,  Xin {\it et al.}~\cite{Xin.19}.

%%Na spectrum simulation with RKKY model
In order to investigate the origin of the spectral components, we invoke a RKKY model describing a long range magnetic coupling between  local moments on the Cu and Fe sites. Within this picture, the magnetic $\mathrm{Cu}^{2+}$ impurities are polarized by the external field, $\textbf{\textit{H}}_{0}$, giving rise to local fields, which induce a spin polarization, \textbf{\textit{s}}(\textbf{\textit{r}}), of the conduction electrons at position specified by $\textbf{\textit{r}}$; $\textbf{\textit{s}}(\textbf{\textit{r}})\propto \chi_\mathrm{RKKY}(\textbf{\textit{r}}) \textbf{\textit{H}}_{0}$, with the magnetic susceptibility given by ~\cite{Rud.54,Boy.74,Boy.76}:

\begin{equation}
\label{Eq.1}
\begin{split}
\chi_{\textrm{RKKY}}(\textbf{\textit{r}}) \propto \sum_{i}\frac{g^2\mu_B^{2}}{2 k_\mathrm{F}}\frac{\textrm{sin}\left[2k_\mathrm{F}(\textbf{\textit{r}}-\textbf{\textit{r}}_{\textrm{Cu},i})+\phi\right]}{|\textbf{\textit{r}}-\textbf{\textit{r}}_{\textrm{Cu},i}|^4}\\-g^2\mu_B^2\frac{\textrm{cos}\left[2k_\mathrm{F}(\textbf{\textit{r}}-\textbf{\textit{r}}_{\textrm{Cu},i})+\phi\right]}{|\textbf{\textit{r}}-\textbf{\textit{r}}_{\textrm{Cu},i}|^3}
\end{split}
\end{equation}

\noindent where $k_{F}$  is the Fermi wave vector and $\phi$ is the average phase shift due to disorder of the dopants~\cite{Jag.88}, and $\textbf{\textit{r}}_{\mathrm{Cu},i}$ is
the location of the $i^{th}$ Cu ion, for which the summation over $i$ represents contributions to the susceptibility at \textbf{\textit{r}} from all Cu ions in the conduction plane. The spin-polarized conduction electrons are coupled through exchange interaction to the local moments at the Fe sites, giving rise to a transferred hyperfine field, \textbf{\textit{h\,$_\mathrm{RKKY}$}}, at their near-neighbor Na nuclei, 
\begin{equation}
	\label{Eq.P3RKKYfield}
	\frac{\textit{h\,$_\mathrm{RKKY}$}}{H_{0}}  = A_{cc}\sum_{j = \textit{NN}}\chi_{\textrm{RKKY},j}, 
\end{equation}
\noindent where $A_{cc}$ is the transferred hyperfine field form factor given by one of the diagonal components of the the hyperfine coupling tensor (Appendix A). This constant is determined by the slopes of $K(T)$ versus $\chi(T)$ for each $^{23}$Na spectral component~\cite{Kit.08}; inset Fig.~\ref{Fig.1}\,(b), where $K(T)$ is the frequency shift determined by $K(T) = \triangle f /f_{L} = (f(T) - f_{L})/f_{L}$, and $\chi(T)$ is the bulk susceptibility. Thus the hyperfine field at any Na site is a sum of the hyperfine coupling to its nearest-neighbor (NN) Fe atoms, at $\textbf{\textit{r}}_{j}$ ($j$ = 1-4). A detailed understanding of the origin of the transferred hyperfine field at the Na sites is beyond the scope of the present work. 
%and would require further investigation in the future.}
To verify this RKKY model, we performed a computer-simulation of the room-temperature $^{23}$Na spectrum shown for $x = 0.012$; the simulated spectrum is essentially a histogram of the hyperfine fields at the Na sites.

However, Eq.\,\ref{Eq.P3RKKYfield} only addresses the magnetic coupling in the spin degrees of freedom. The frequency shift measured for each $^{23}$Na spectral component, shown in Fig.~\ref{Fig.1}\,(b) must be corrected for the temperature independent contribution from the chemical shift, $K_c$, given  by linear extrapolation of  $K(\chi)$ to zero (inset Fig.~\ref{Fig.1}\,(b)), for each spectral component. The $K_{c}$-corrected Gaussian fits to the experimentalq spectrum, can then be compared with the simulation shown in Fig.~\ref{Fig.1}\,(c). The simulation was performed with a 800~$\times$~800 Fe lattice, randomly populated by Cu dopants with a probability $p = 0.012$. A small portion of the simulated Fe-Cu lattice is depicted in Fig.~\ref{Fig.1}\,(e), showing a broader view of \textbf{\textit{h}}$_{\textrm{RKKY}}$ at each Na site. Five concentric shells of Na sites, centered at a single Cu dopant are identified [Fig.~\ref{Fig.1}\,(d)], associated with the spectral components labeled from smallest to largest in Fig.~\ref{Fig.1}\,(c). Each shell represents a group of Na nuclei contributing to one of the spectral components. The radius of each shell is determined by the average distance of the corresponding Na sites to the Cu-dopant in the $ab$-plane, with the exception of the 5$^{\textrm{th}}$ shell which corresponds to all Na sites contributing to the main spectral component. The dependence of the simulated $\chi_{\textrm{RKKY}}$ on distance, $r$, from a single impurity at the origin, is shown in Fig.~\ref{Fig.1}\,(f), revealing a damped oscillation between ferromagnetic and antiferromagnetic coupling. At short distance, $r \lesssim d$, where $d$ is the lattice spacing between two NN Fe atoms, the RKKY interaction is ferromagnetic. In the same figure, we compare the average $h_{\textrm{RKKY}}$ associated with each spectral component between the simulated and experimental spectra corrected for $K_{c}$.  The reasonable agreement is within experimental uncertainty. We also compare the relative spectral weight associated with each spectral component between the two, as shown in the inset of Fig.~\ref{Fig.1}\,(f). This comparison shows a good agreement with the exception of the spectral component color coded green in Fig.~\ref{Fig.1}\,(b) and (c). We note the possible existence of additional Na sites contributing to this particular spectral component, which are not considered in our model. Indeed, our assumption of RKKY interaction in the form of Eq.~\ref{Eq.1} could break down in small regions of the lattice where more than one dopants are located relatively close to one another. Incidentally, these regions are occupied by Na sites that are colored coded black in Fig.~\ref{Fig.1}\,(e), contributing to none of the spectral components based on our model. Overall, our model provides a good representation of the field distribution at Na nuclei due to the RKKY interaction among the local moments at the Fe and Cu sites.  Direct hyperfine coupling between Na and conduction electrons is neglected. Contributions to the hyperfine fields at Na sites from the next-nearest Fe-Cu layer can also be ignored.
%for that the transferred hyperfine interaction is usually short-range.}

\begin{comment}
Different contributions to $K(T)$ can be written as $K(T) = K_{s}(T) + K_{o}+K_{q}$, where $K_s(T)$ is the temperature dependent spin shift, $K_{o}$ is the orbital shift, and $K_{q}$ is the quadrupolar shift which effects the central transition only through second-order and is expected to be negligible due to symmetry of the \textit{NN} Fe atoms for a given Na nuclei. Since Eq.\,\ref{Eq.P3RKKYfield} underlines the magnetic coupling in the spin degree of freedom only, the orbital contribution, $K_{o}$, to the frequency shift, $K$, must be subtracted from each spectral component, inset Fig.~\ref{Fig.1}\,(b), before the simulation results can be compared with the experimental spectrum. 
\end{comment}

For the fit of the NMR spectrum to the RKKY model, we chose $k_{\mathrm{F}} \approx 0.18\pi/d$. This value is reasonable for iron pnictides with a Fermi surface composed of both hole and electron pockets~\cite{Akb.13}. The average phase shift owing to the random distribution of the dopants was found to be $\phi \approx 0.8\pi$.  

In addition to the RKKY model, one might attribute the $^{23}$Na NMR satellites to inequivalent Na sites, each of which is associated with a different number of nearest-neighbor Fe sites occupied by a Cu dopant, similar to that discussed in Ref.~\cite{Xin.19} for the more insulating compounds $x > 0.1$. However, if this were the case, the scarcity of these sites due to low Cu concentration would lead to satellites of spectral weight that is at least an order of magnitude smaller than what has been observed from the NMR spectra which we report here.

%TODO: the following should be discussed in the appendix
\begin{comment}
In addition to the RKKY interaction, one needs to note that the observed spectral weight of different NMR peaks might have been due to the samples having a significantly higher Cu concentration than $\sim1\%$. However, this possibility is ruled out as the measurement of the bulk susceptibility unambiguously shows a superconducting transition around $T \approx 10$ K, indicating that the actual Cu concentration is indeed around 1\%~\cite{Wan.13}.
\end{comment}

\section{Spin-Lattice Relaxation}

\begin{figure*}
	\hspace{-0.8cm}
	\includegraphics[scale = 0.14]{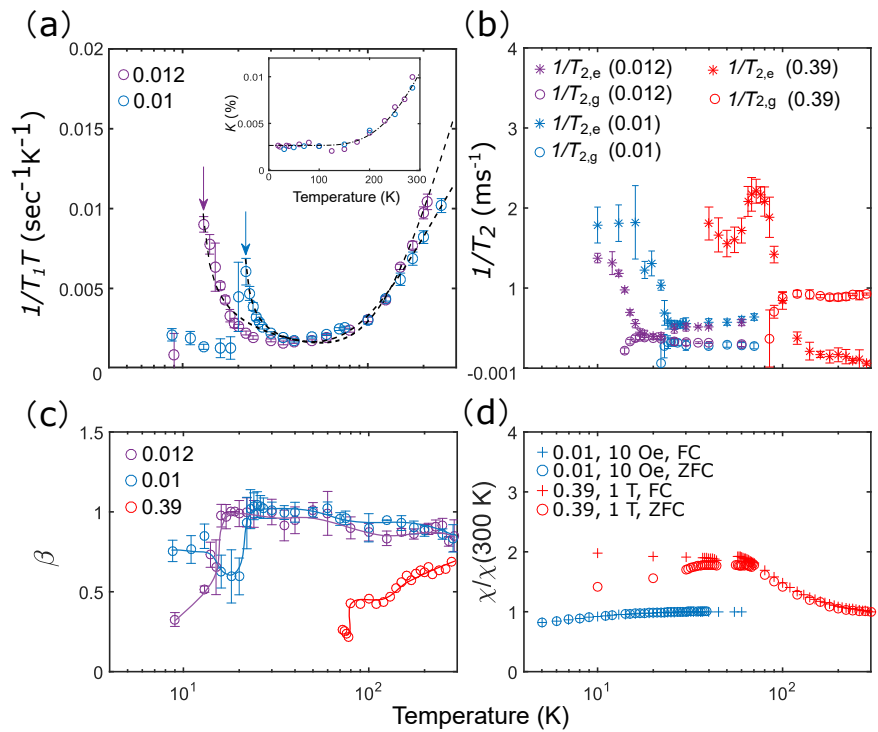}
	\caption{\label{Fig.2} Magnetic transitions. (a) Spin-lattice relaxation ($1/T_{1}T$) measurement measured for the central transition of the main spectral component in $x = 0.012$ and $0.01$. The dashed and solid lines are fits that capture both a pseudogap and a Curie-Weiss-like behavior of the data. The fits give $T_{\textrm{AF}}\approx 11$ K and 20 K for $x$ = 0.012 and 0.01 respectively, marked by arrows. At high temperatures both  $K(T)$ (inset) and $1/T_{1}T$ are due to the pseudogap.(b) Gaussian and exponential components of the spin-spin relaxation rate ($1/T_{2}$). (c) Stretched exponent $\beta$. (d) Bulk magnetic susceptibility for $x = 0.010$ and 0.39, with $H = 10$ Oe and 1 T$~||~ab$. The bifurcation between data field-cooled (FC) and zero-field-cooled (ZFC) conditions for $x = 0.39$ demonstrates a classic manifestation of spin-glass, absent for $x \approx 0.01$.}
\end{figure*}

Spin-lattice relaxation measurements can be very informative regarding the nature of  AFM order in pnictide compounds as demonstrated by Ning {\it et al.}~\cite{Nin.10},  Dioguardi {\it et al.}~\cite{Dio.10,Dio.13}, and Oh~\cite{Oh.12} in their $^{75}$As studies of the Ba122 system and with $^{23}$Na NMR in Co doped Na111 compounds by Oh {\it et al.}~\cite{Oh.13}.  Our data for $1/T_{1}T$ of the main spectral component in both $x = 0.010$ and 0.012, for $H_0 ||c-$axis, is presented in  Fig.~\ref{Fig.2}\,(a). These measurements  probe spin dynamics at nonzero wave vector $\textbf{\textit{q}}$ and are related to the imaginary part of the $\textbf{\textit{q}}$-dependent dynamic susceptibility $\chi^{''}$: $1/T_1T \sim \sum\limits_{\textbf{\textit{q}}}|A(\textbf{\textit{q}})|^2\,\chi^{''}(\textbf{\textit{q}},f_L)/f_L$, where $A(\textbf{\textit{q}})$ is the form factor of the hyperfine interaction between the electronic and nuclear spins. The $1/T_1T$ versus $T$ data were fit to a combination of pseudogap and Curie-Weiss-like terms, $1/T_{1}T = (1/T_1T)_{\textrm{pg}} + (1/T_1T)_{\textrm{AF}} = [A ~\textrm{exp}(-\triangle_{pg}/k_BT) + B] + C / (T - T_{\mathrm{AF}})^\gamma$, where the critical exponent is $\gamma = 0.5$. This fit has been used previously to account for contributions to the spin-lattice relaxation from both pseudogap and AFM fluctuations in pnictides~\cite{Nin.10,Dio.10}. The effects of the pseudogap on $K(T)$ and $1/T_{1}T$, over a range of composition, die out at low temperatures as previously reported~\cite{Xin.19,Xin.20}. While the pseudogap is clearly shown by $1/T_{1}T$ and $K$ at temperatures $T \gtrsim 100$ K, the Gaussian component of the spin-spin relaxation rate, $1/T_{2,g}$, which probes the real part of the static susceptibility $\chi'(\textbf{\textit{q}},\omega = 0)$, is largely temperature-independent in the paramagnetic state; Fig.~\ref{Fig.2}\,(b). This, combined with the decrease in $1/T_{1}T$, suggests an increase in the ratio, $T1T/T_{2g}$, consistent with opening of a spin pseudogap~\cite{Aul.97}.  At temperatures, $T\lesssim 40$ K, $(1/T_{1}T)_{\textrm{AF}}$ becomes dominant and the critical exponent of $\gamma = 0.5$ corresponds to 3D spin fluctuations due to weak itinerant antiferromagnetism~\cite{Mor.74} at the AFM wave vector \textbf{\textit{q}}$_{\textrm{AF}}$ = ($\pi/d$,0)~\cite{Tan.17}.  This interpretation is consistent with the observed temperature independent behavior of $\chi(\textbf{\textit{q}} = 0)$ probed by $K(T)$ in the same temperature range, inset Fig.\ref{Fig.2}\,(a). In contrast, the critical exponent clearly deviates from $0.5$ for compounds in the heavily-doped regime. Setting $\gamma$ as a free parameter for fitting, we obtain for comparison,  $\gamma \approx 1.5 \pm 0.1$ and $1.57\pm 1.30$ for $x=0.39$ and 0.48, respectively (Appendix C). We infer that the magnetic state evolves from  itinerant to  localized as $x$ exceeds $\sim 0.3$~\cite{Cha.17}.

\begin{comment}
This clear deviation from $\gamma = 0.5$ signifies a qualitative difference corresponding to a change in the nature of the magnetism with $x$ increasing beyond $\sim0.2$, consistent with the observed metal-to-insulator transition in the doping regime $x \gtrsim 0.3 $~\cite{Cha.17}. 
\end{comment}

%%%%%%%%%%%%%%%%%   F I G U R E  2   %%%%%%%%%%%%%%%%%%

%%%%%%%%%%%%%%%%%%%%%%%%%%%%%%%%%%%
Despite these differences, we note that the magnetic transition in $x = 0.010$ and 0.012 share some similarity with that in $x \gtrsim 0.3 $: (i) the magnetization recovery from the spin-lattice relaxation exhibits a stretched exponential behavior (Appendix B) with $T$ approaching the AFM transition temperature $T_{\textrm{AF}}$~[Fig.~\ref{Fig.2}\,(c)], indicating a distribution of relaxation rates, (ii) the fraction of $^{23}$Na nuclei contributing to the NMR spectrum, $N_{0}$, diminishes with decreasing temperature, and (iii) the spin-spin relaxation rate $1/T_{2}$ can be separated into Gaussian $T_{2,g}$ and the exponential $T_{2,e}$ components (Appendix B).  A crossover from the former to the latter occurs as the temperature approaches $T_\textrm{AF}$~[Fig.~\ref{Fig.2}\,(b)]. These observations are strong indications of inhomogeneous local magnetic fields, and is a general signature of glassy magnetic phases at low temperatures~\cite{Cur.00,Dio.13}. This is also the case for large $x = 0.39$~\cite{Xin.20}. However, it is unlikely that the inhomogeneity of local fields for $x = 0.010$ and 0.012 is due to magnetic frustration in contrast with $x = 0.39$. We note that a majority of the Na nuclei that contribute to the main spectral component are sufficiently far from Cu dopants that  near neighbor Fe atoms cannot be frustrated due to competing RKKY interactions. Furthermore, the independence of Cu dopants in $x = 0.010$ and $0.012$ precludes significant structural disorder leading to magnetic frustration as is the case for $x = 0.39$~\cite{Xin.20}.  This inference is also consistent with the fact that the stretched exponent $\beta$ for $x = 0.010$ and $0.012$ only starts to decrease from $\beta = 1$  close to $T_{\textrm{AF}}$, Fig.~\ref{Fig.2}\,(c). In contrast, $\beta$ for $x = 0.39$ is clearly already suppressed at room temperature, indicating intrinsic magnetic inhomogeneity and frustration due to the aforementioned structural disorder caused by dopants. The time recovery of both the spin-lattice and spin-spin relaxation is discussed in Appendix B. No obvious bifurcation has been observed between the bulk susceptibility $\chi(T)$ taken under zero-field-cooled (ZFC) and field-cooled (FC) conditions for $x = 0.010$, even for the small applied field $H = 10$ Oe, as shown in Fig.~\ref{Fig.2}\,(d). This offers further evidence of negligible frustration effects in $x = 0.010$ and 0.012.
%%%%%%%%%%%%%%%%%%%%%%%%%%%%%%%%%%%%%%%%%%%%%%%%%%%%%%%%%%%%%%%%%%
\section{Magnetic Inhomogeneity}
%in which the magnetic transition is spin-glass in nature that is attributable to the frustrated Fe-Cu square lattice.

%% computer simulation Cu perturbation on the AFM background
The $^{23}$Na linewidth associated with all satellite spectral components increases substantially as the temperature approaches the AFM transition (Appendix D), rendering the satellites difficult to be resolved. To determine the origin of the magnetic inhomogeneity in $x = 0.010$ and 0.012, we compared our spectra for the central transitions with a simulation for the distribution of local fields. We adopted a Gaussian model to describe the AFM background in which the magnetic moments at Fe sites, $\textbf{\textit{m}}_{\textrm{AF}}(\textbf{\textit{r}}_{\textrm{Fe}})$, point along the crystalline $a$-axis and are suppressed in the vicinity of Cu dopants~\cite{Dio.10},

\begin{equation}
\label{Eq.2}
\begin{split}
&\textbf{\textit{m}}_{\textrm{AF}}(\textbf{\textit{r}}_{\textrm{Fe}}) = m_\textrm{AF}(\textbf{\textit{r}}_{\textrm{Fe}})\hat{\textbf{\textit{a}}} 
= C_0~\textrm{cos}(\textbf{\textit{q}}_{\textrm{AF}}\cdot \textbf{\textit{r}}_\textrm{Fe})\\
&\times\{1-C_1\sum_{i}\textrm{exp}[-|\textbf{\textit{r}}_{\textrm{Fe}}-\textbf{\textit{r}}_{\textrm{Cu},i}|^{2}/2\xi^{2}]\}\hat{\textbf{\textit{a}}}\\
&=C_0(-1)^{n_a}[1-C_1\sum_{i}\textrm{exp}(-|\textbf{\textit{r}}_{\textrm{Fe}}-\textbf{\textit{r}}_{\textrm{Cu},i}|^{2}/2\xi^{2})]\hat{\textbf{\textit{a}}}
\end{split}
\end{equation}

\noindent where $\textbf{\textit{q}}_{AF} = (\pi/d,0)$~\cite{Tan.17}, and $\xi$ the length scale of this suppression. The constant $C_{0}$ represents the magnitude of the AFM ordered moment and $C_{1}$ the suppression. Thus the magnitude of the total hyperfine field, $h_{\textrm{total}}$, at a Na nucleus can be approxiamted by,

\begin{equation}
	\label{Eq.3}
	\begin{split}
	h_{\textrm{total}} \approx \pm A_{ca}\sum_{j=\mathrm{NN}}(-1)^{j}m_{\textrm{AF}}(\textbf{\textit{r}}_{\textrm{Fe},j})\\ + A_{cc}H_{0}\sum_{j = \mathrm{NN}}\chi_{\mathrm{RKKY}}(\textbf{\textit{r}}_{\textrm{Fe},j}) 
	\end{split}
\end{equation}
\noindent where $A_{ca}$ is the off-diagonal hyperfine field form factor (Appendix A) approximated to be $A_{ca} \approx 0.027 ~ \textrm{T}/ \mu_{B}$~\cite{Ma.11}, the plus and minus signs in front of $A_{ca}$ correspond to  Na nuclei above and below the Fe plane~\cite{Wan.13}. The susceptibility $\chi_{\mathrm{RKKY}}$ is given by the RKKY simulation based on Eq.\,\ref{Eq.1} for the room-temperature spectrum. We fit the main component of the $^{23}$Na NMR spectrum 
%\red{within its full width at half maximum (FWHM)} 
of $x = 0.012$ at $T=9$ K to the histogram of $h_\textrm{total}$ given by Eq.\,\ref{Eq.3} with results in Fig.~\ref{Fig.3}\,(a).  %The experimental and simulated spectra are arbitrarily aligned at zero frequency. 
From the fit we obtained $C_0\approx1.3$, $C_1\approx0.64$, and $\xi\approx6.5d$, indicating that the Cu dopant in NaFeAs gives rise to a magnetic perturbation that has similar magnitude but a longer range than Ni does in the Ba122 system~\cite{Dio.10}, possibly due to the fact that the Cu atoms are associated with a stronger magnetic scattering potential than impurities in Ba122~\cite{Kem.09,Wad.10}. A lorentzian model gives similar results, also shown in Fig.~\ref{Fig.3}\,(a). The Cu-induced magnetic perturbation, increases with $x$ in the doping range $ 0<x<0.02$, consistent with the suppression of the AFM order shown in the phase diagram (Fig.~\ref{PhaseDiagram}). The exchange interaction between Cu and Fe sites could give rise to ordered moments at the former, leading to additional contributions to $\textbf{\textit{h}}_\textrm{RKKY}$ at the Na sites in the AFM phase. However, such contribution can be neglected in our simulation as we mainly focus on the broadening of the main peak due to the Cu-induced suppression of the background AFM order. The Na nuclei associated with this peak are located sufficiently far away from the Cu dopants so that a change in $\textbf{\textit{h}}_\textrm{RKKY}$ would have negligible effect on these nuclei.

\begin{comment}
Since $A_{cc}\propto\langle|\psi_{\textbf{k}}(0)|^2\rangle_{E_{\textrm{F}}}$, the modulus square of the electron wave function at the $^{23}$Na nuclei integrated over the Fermi surface, its $r$ dependence may originate from the Cu-induced perturbation on the local charge density that the extra $3d$ electrons doped by Cu are trapped around the dopant~\cite{Wad.10,Ide.13}. By fitting $A_{cc}$ versus $r/d$ to the model $A_{cc}(r/d) = a~\textrm{exp}[-(r/d)^2/(2\lambda^2)]+b$, we obtained $\lambda \approx d_{\textrm{Fe-Fe}}$ which is significantly smaller than $\xi \approx 6.5d$. This is a possible indication that the Cu-induced perturbation in the magnetic channel extends much further than in the charge channel~\cite{Kem.09}, owing to that the length scale of the magnetic perturbation could be enhanced by the background long-range AFM order, which has a correlation length $\xi_{\textrm{AF}} \gtrsim 35d$~\cite{Wan.18}.
\end{comment}
 
The RKKY interaction is particularly sensitive to the electronic structure of the system. In the paramagnetic state, the frequency shift associated with each satellite appears to be following approximately the same temperature dependence down to $\sim125 ~\textrm{K}$ (Appendix D), suggesting that the RKKY interaction is largely temperature-independent at high temperatures. However, the amplitude of the RKKY oscillation could weaken due to a spin-density-wave (SDW) gap in the AFM state~\cite{Akb.11, Yi.12}, leading to smaller hyperfine fields at the Na sites and therefore smaller frequency shifts of the satellites with respect to the main peak. This change in relative frequency shift, combined with the line broadening of the NMR spectrum, could lead to the satellite resonances being absorbed into the main peak, and therefore rendering the satellites absent in the AFM state, as shown in the experimental spectrum at $T = 9~\textrm{K}$. This is in contrast to the simulated spectra for which the effect of SDW gap was not considered [Fig.~\ref{Fig.3}\,(a)]. A scanning tunneling microscopic study of the SDW gap in the parent compound is discussed in Ref.~\cite{Zho.12}.

\section{Coexistence of local and itinerant magnetism}
From the simulation results we approximated the ordered moment of the AFM background to be $m\approx0.003$ $\mu_{B}/\textrm{Fe}$, smaller than that approximated from the linear interpolation of the neutron scattering results for $x = 0.016$ and the parent compound, $m \approx 0.02$ $\mu_{B}/\textrm{Fe}$~\cite{Son.16}. This discrepancy could be partly attributed to the relatively large uncertainty associated with the moment for the parent compound. Also, the evolution of the ordered moment with doping may not be linear, as reflected for the Co-doped Na-111 system~\cite{Tan.16}. More importantly, thermal fluctuations could suppress the ordered moment probed by NMR which has a time scale  at least three orders of magnitude longer than that of the neutron scattering. A NMR study extending into the superconducting state can reveal further information about the evolution of the AFM order moment at lower temperatures $T \leq T_{c}$.  However, this is beyond the scope of the present article. Despite the discrepancy between the NMR and neutron scattering results, the magnetic moments given by the two measurements are both small; orders of magnitude weaker than that for a NaFe$_{0.56}$Cu$_{0.44}$As single crystal ($m \approx 1 ~\mu_{B}/\textrm{Fe}$ at $T \approx 4 \,\textrm{K}$)~\cite{Son.16}. Such small magnetic moments in $x\approx0.01$ 
are aligned with itinerant antiferromagnetism, consistent with our analysis of the spin-lattice relaxation data. Another possible cause for the small ordered moment could be the presence of magnetic frustration~\cite{Si.08}; however, we found no evidence for significant frustration from inhomogeneous relaxation, from our simulation, or from $\chi(T)$.

%%%%%%%%%%%%%%%%%   F I G U R E  3   %%%%%%%%%%%%%%%%%%
\begin{figure}
	\hspace*{-0.22cm}
	\includegraphics[scale = 0.14]{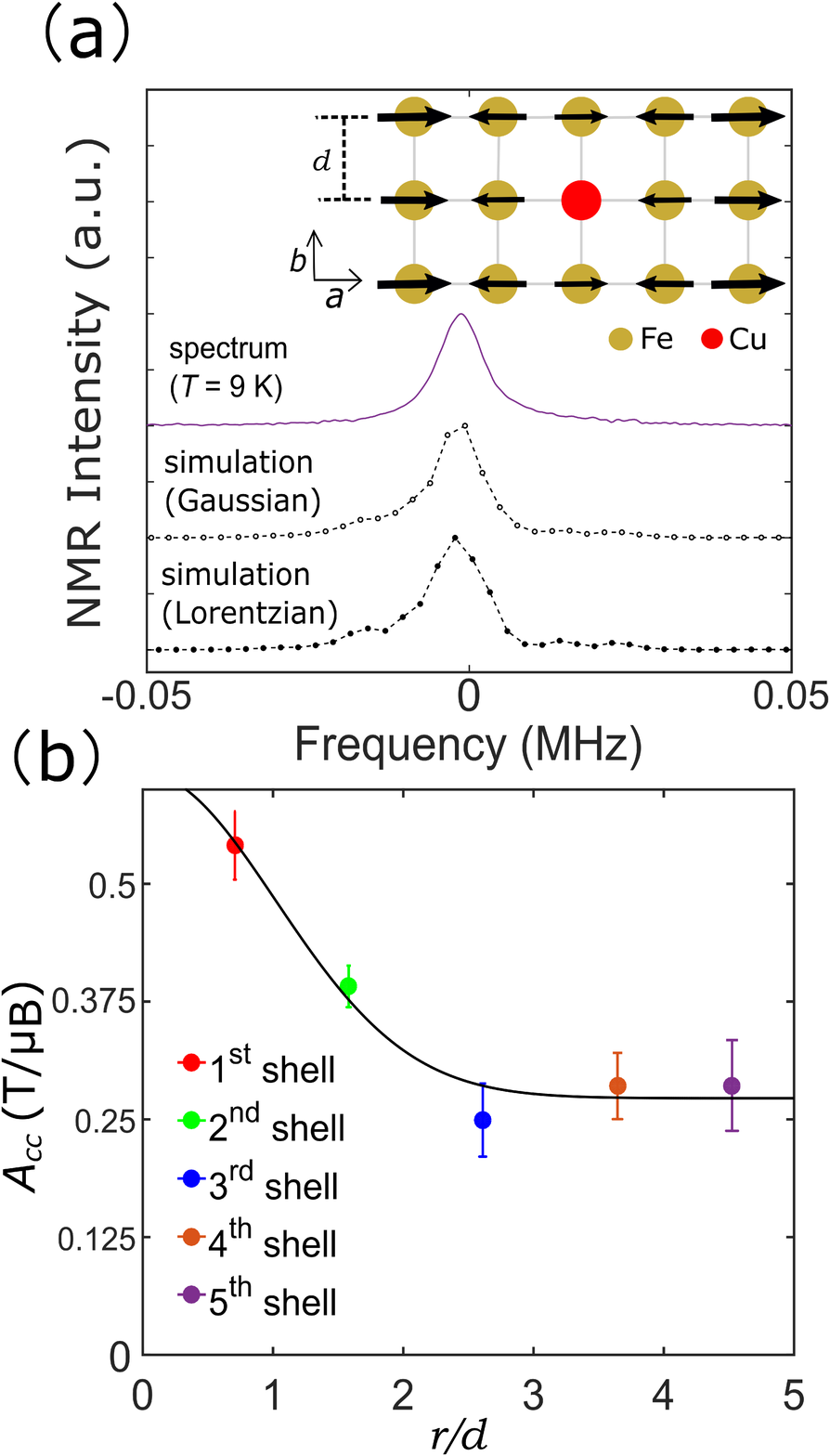}
	\caption{\label{Fig.3} Analysis of $^{23}$Na NMR line shapes from impurity induced perturbation of the AFM background. (a) Comparison between the $^{23}$Na spectrum for $T = 9$ K and $H_{0} = 14~\textrm{T}~||~c$-axis, and the simulated spectra. Both Gaussian and  lorentzian models were considered leading to similar length scales $\xi\approx6.5\,d$, where $d$ is the Fe lattice spacing. A schematic of the magnetic perturbation is shown in the inset; black arrows represent perturbed magnetic moments. (b) Dependence of the hyperfine field form factor, $A_{cc}$, on $r/d$. The solid curve represents a fit to our model for $A_{cc}(r/d)$.}
\end{figure}  

%%%%%%%%%%%%%%%%% %%%%%%%%%%%%%%%%%%

The conduction electrons are mainly responsible for the long-range magnetic order; however, the existence of localized moments is intrinsic to the  RKKY model since the local moments are coupled through that interaction. Using the RKKY simulation results and $K(\chi)$, we calculate the $r$-dependence of the hyperfine field form factor, $A_{cc}$ vs $r/d$, in Fig.~\ref{Fig.3}\,(b), with $r$ being the distance from a Cu dopant in the $ab$-plane. Since $A_{cc}$ characterizes the strength of the transferred hyperfine interaction between the Fe $3d$ electrons and the Na nuclei, its $r$-dependence may originate from a change in the local electron density around a Cu dopant~\cite{Wad.10,Ide.13}. Indeed, for iron pnictides, it was demonstrated that the extra $d$ electrons contributed by the dopants are largely localized in their near vicinity, leading to a strong increase in the electron density close to the dopant~\cite{Wad.10}. Fitting the data to the phenomenological model, $a\,\textrm{exp}[-(r/d)^2/(2\lambda^2)]+b$ [Fig.~\ref{Fig.3}\,(b)], we obtain $\lambda \approx d \pm0.8d$, significantly smaller than the correlation length of the Cu-induced perturbation to the AFM order; $\xi \approx 6.5\,d$. The result of this comparison can be attributed to possible enhancement of the magnetic perturbation by the background long-range AFM order, which has a correlation length $\xi_{\textrm{AF}} \gtrsim 35\,d$~\cite{Wan.18}. On the other hand, in a metallic crystal, an excessive impurity potential responsible for a perturbation to the local electron density can be  almost completely screened by conduction electrons within the impurity Wigner-Seitz cell~\cite{Gal.19}. Thus, the relative small value of $\lambda$ might also be ascribed to the screening effect of the conduction electrons.

Our direct evidence for the RKKY interaction, combined with analysis of the spin-lattice relaxation rate and the Cu-induced magnetic perturbation, indicate that the picture of coexisting itinerant and localized electrons is important for understanding the magnetism in lightly Cu-doped \NaFeCuAs. This is consistent with theoretical studies in pnictides showing that part of the Fe $3d$ electrons are itinerant and the rest are localized~\cite{Yin.10, Wu.08, Med.09}. Moreover, neutron scattering results show that the Bragg peak remains at a commensurate position in compounds approaching optimal doping ($x\approx 0.02$)~\cite{Tan.17}, which is not compatible with a purely itinerant picture where Fermi surface nesting gives rise to incommensurate magnetic order. The existence of RKKY interaction also reveals commonality between metallic pnictides and other strongly correlated electron systems including heavy fermion compounds, prompting future identification of quantum criticality and study of its nature in Cu-doped \NaFeCuAs.

\section{Conclusion}
In summary, using $^{23}$Na NMR measurements on the lightly Cu-doped \NaFeCuAs ~($x\approx0.01$) single crystals and numerical simulation, we have shown direct evidence of RKKY interaction between the local moments at the Cu and Fe sites in this pnictide system. For the AFM phase, we have shown that the magnetic order exhibits itinerant nature, associated with a small ordered moment. The ordered moments at the  Fe sites are perturbed in the vicinity of the Cu dopants, giving rise to magnetic inhomogeneity. Our NMR results indicate coexistence of local and itinerant magnetism in the lightly Cu-doped \NaFeCuAs. 

%For iron chalcogenides and heavy-fermion materials, the Hund's rule coupling between the local and itinerant electrons induces RKKY-like exchange and give rises to long-range magnetic order~\cite{Hir.15, Isa.13}. Since previous theoretical calculations based on a double-exchange model have shown that the Hund's rule coupling also plays a crucial role in determining the type of AFM order in pnictides system~\cite{Yin.10}, our NMR results extend naturally from this study and provide  experimental evidence for a common ground shared by iron pnictides and other families of strongly-correlated material.}

% Similarly in pnictides, it was shown that the Hund's rule coupling plays a crucial role in determining the type of AFM order~\cite{Yin.10}.

%= a~\textrm{exp}[-(r/d_{\textrm{Fe-Fe}})^2/(2\lambda^2)]+b$, giving $\lambda\approx d\pm0.8d$. 

\section{acknowledgment}
We thank Weiyi Wang and Chongde Cao for their contributions to the crystal growth and characterization. The NMR spectrometer, MagRes2000  wide-band spectrometer system, was designed by A. P. Reyes at the National High Magnetic Field Laboratory (NHMFL). The home-built continuous flow cryostat was designed by J.A. Lee at Northwestern University. Research was supported by the U.S. Department of Energy (DOE), Office of Basic Energy Sciences (BES), Division of Material Sciences and Engineering under Award No. DE-FG02-05ER46248 (WPH) and DE-FG02-05ER46202 (PD), and the NHMFL by NSF and the State of Florida. The single crystal growth efforts at Rice were supported by the U.S. DOE, BES under Grant No. DE-SC0012311.  Part of the materials work at Rice was supported by the Robert A. Welch Foundation under Grant No. C-1839.

%%%%%%%%%%%%%%%%%%%%%%%%%%%% Beging of Appendix%%%%%%%%%%%%%%%%%%%%%%%%%%%%%%%%%%%%
\newcommand{\beginappendix}{%
	\setcounter{table}{0}
	\renewcommand{\thetable}{A\arabic{table}}%
	\setcounter{figure}{0}
	\renewcommand{\thefigure}{A\arabic{figure}}%
	\setcounter{equation}{0}
	\renewcommand{\theequation}{A\arabic{equation}}%
}

\beginappendix
\subsection{APPENDIX A: Transferred hyperfine fields at Na Sites}
\begin{figure}[h]
	\includegraphics[scale=0.45]{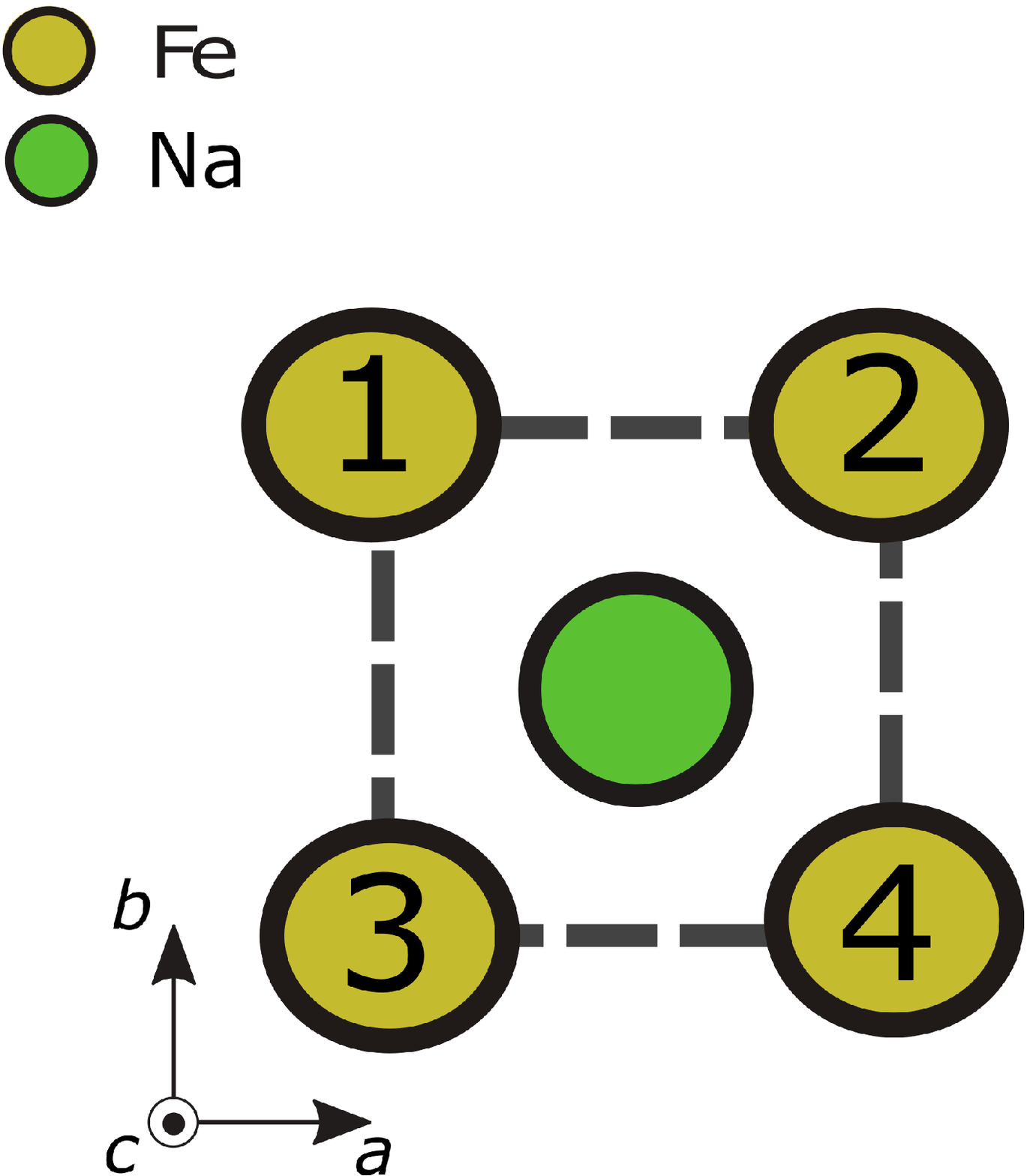}
	\caption{Schematic of a Na nucleus and its four nearest-neighbor (NN) Fe$^{2+}$ ions. Each Fe$^{2+}$ ion is labeled and associated with one hyperfine coupling tensor; Eq.~\ref{Eq.HFtensor}.}\label{SFig_Feconfiguration}
\end{figure} 

The coupling of a Na nucleus to on-site orbitals and orbitals on surrounding Fe$^{2+}$ ions gives rise to transferred hyperfine field at the Na site. For the Fe-Na configuration in Fig.~\ref{SFig_Feconfiguration}, the transferred hyperfine coupling tensors, $\tilde{A}_{i}$ ($i = 1-4$), are defined by:
% hypefine coupling tensor
\begin{equation}
	\label{Eq.HFtensor}
	\begin{split}
		\tilde{\mathbf{A}}_1 = 
		&\begin{pmatrix}
			A_{aa} & -A_{ab} & -A_{ac} \\ 
			-A_{ba} & A_{bb} & A_{bc} \\
			-A_{ca} & A_{cb} & A_{cc}
		\end{pmatrix},\\
		\tilde{\mathbf{A}}_2 =& 
		\begin{pmatrix}
			A_{aa} & A_{ab} & A_{ac} \\ 
			A_{ba} & A_{bb} & A_{bc} \\
			A_{ca} & A_{cb} & A_{cc}
		\end{pmatrix},\\
		\tilde{\mathbf{A}}_3 = 
		&\begin{pmatrix}
			A_{aa} & A_{ab} & -A_{ac} \\ 
			A_{ba} & A_{bb} & -A_{bc} \\
			-A_{ca} & -A_{cb} & A_{cc}
		\end{pmatrix},
		\\\hspace{2pt} \tilde{\mathbf{A}}_4 = &
		\begin{pmatrix}
			A_{aa} & -A_{ab} & A_{ac} \\ 
			-A_{ba} & A_{bb} & -A_{bc} \\
			A_{ca} & -A_{cb} & A_{cc}
		\end{pmatrix},	
	\end{split}
\end{equation}
\noindent each of which is associated with the coupling of the Na nucleus with one of its nearest-neighbor (NN) Fe$^{2+}$ ion. In the paramagnetic state with $\textbf{\textit{H}}_{0} \parallel c$-axis, the magnetic moment ($\textbf{\textit{m}}_{i}$) of a Fe$^{2+}$ ion is polarized along the $c$-axis; $\textbf{\textit{m}}_i = \chi_{i,\textrm{RKKY}}\textbf{\textit{H}}_0$. The susceptibility due to the RKKY interaction, $\chi_{i,\textrm{RKKY}}$, is given by Eq.~1 in the main text. Therefore, the hyperfine field at a Na site is given by

\begin{equation}
	\label{Eq.hyperfinefield}
	\begin{split}
		\textbf{\textit{h}}_\textrm{RKKY} &= \sum_{\textrm{NN}} \mathbf{\tilde{A}}_{i}\cdot \textbf{\textit{m}}_{\textrm{RKKY},i} 
		= H_0\sum_{\textrm{NN}} \mathbf{\tilde{A}}_{i}\cdot \begin{pmatrix}
			0\\
			0\\
			\chi_{\textrm{RKKY},i}
		\end{pmatrix}\\
		&=\biggl[\chi_{\textrm{RKKY},1}\begin{pmatrix}
			-A_{ac}\\
			A_{bc}\\
			A_{cc}
		\end{pmatrix} + 
		\chi_{\textrm{RKKY},2}\begin{pmatrix}
			A_{ac}\\
			A_{bc}\\
			A_{cc}
		\end{pmatrix}\\&+\chi_{\textrm{RKKY},3}\begin{pmatrix}
			-A_{ac}\\
			-A_{bc}\\
			A_{cc}
		\end{pmatrix}+
		\chi_{\textrm{RKKY},4}\begin{pmatrix}
			A_{ac}\\
			-A_{bc}\\
			A_{cc}
		\end{pmatrix}\biggl]H_0
	\end{split}
\end{equation}

\noindent In the paramagnetic state,  the total field can then be approximated by its component along the $c$-axis,  $\textbf{\textit{H}}_\textrm{total} = \textbf{\textit{H}}_{0} + \textbf{\textit{h}}_{\textrm{RKKY}} \approx \textbf{\textit{H}}_{0} + A_{cc}\textbf{\textit{H}}_{0}\sum_{\textrm{NN}}\chi_{\textrm{RKKY},i}$. 

In the antiferromagnetic (AFM) state, the collinear magnetic ordered moments point along the $a-$axis. Therefore for the same Fe-Na configuration, the hyperfine field due to the AFM order can be calculated as 

%tensor calculation for AF component of the hyperfine field 
\begin{equation}
	\label{P3HFfield_AF}
	\begin{split}
		&\textbf{\textit{h}}_\textrm{AF} = \sum_{\mathrm{NN}} \mathbf{\tilde{A}}_{i}\cdot \textbf{\textit{m}}_{\textrm{AF},i}\\
		&= \begin{pmatrix}
			A_{aa}&-A_{ab}&-A_{ac}\\
			-A_{ba}&A_{bb}&A_{bc}\\
			-A_{ca}&A_{cb}&A_{cc}
		\end{pmatrix}\cdot\textbf{\textit{m}}_{\textrm{AF,1}}\\&+
		\begin{pmatrix}
			A_{aa}&A_{ab}&A_{ac}\\
			A_{ba}&A_{bb}&A_{bc}\\
			A_{ca}&A_{cb}&A_{cc}
		\end{pmatrix}\cdot\textbf{\textit{m}}_{\textrm{AF,2}}\\&+
		\begin{pmatrix}
			A_{aa}&A_{ab}&-A_{ac}\\
			A_{ba}&A_{bb}&-A_{bc}\\
			-A_{ca}&-A_{cb}&A_{cc}
		\end{pmatrix}\cdot\textbf{\textit{m}}_{\textrm{AF,3}}\\&+
		\begin{pmatrix}
			A_{aa}&-A_{ab}&A_{ac}\\
			-A_{ba}&A_{bb}&-A_{bc}\\
			A_{ca}&-A_{cb}&A_{cc}
		\end{pmatrix}\cdot\textbf{\textit{m}}_{\textrm{AF,4}}
	\end{split}
\end{equation}
where the AFM ordered moment at the $i^{\textrm{th}}$ NN Fe site is given by $\textbf{\textit{m}}_{\textrm{AF},i} = \textbf{\textit{m}}_{\textrm{AF}}(\textbf{\textit{r}}_{\textrm{Fe},i}) = \textit{m}_\textrm{AF}(\textbf{\textit{r}}_{\textrm{Fe},i})\hat{\textbf{\textit{a}}}$ (Eq.\ref{Eq.2}). Thus,

\begin{equation}
	\label{P3HFfield_AF_2}
	\begin{split}
		&\textbf{\textit{h}}_\mathrm{AF} =
		\begin{pmatrix}
			A_{aa}\\ -A_{ba} \\ -A_{ca}
		\end{pmatrix}m_{\textrm{AF},1} +
		\begin{pmatrix}
			A_{aa}\\A_{ba}\\A_{ca}
		\end{pmatrix}m_{\textrm{AF},2} \\&+
		\begin{pmatrix}
			A_{aa}\\A_{ba}\\-A_{ca}
		\end{pmatrix}m_{\textrm{AF},3} +
		\begin{pmatrix}
			A_{aa}\\-A_{ba}\\A_{ca}
		\end{pmatrix}m_{\textrm{AF},4}
	\end{split}
\end{equation}
\noindent For $\textbf{\textit{H}}_{0} \parallel c$-axis, the total magnetic field in the AFM state can be approximated as $\textbf{\textit{H}}_{\textrm{total}}=\textbf{\textit{H}}_0+\textbf{\textit{h}}_\textrm{RKKY}+\textbf{\textit{h}}_\textrm{AF} \approx \textbf{\textit{H}}_0 + A_{cc}H_0\begin{pmatrix}
	0\\0\\\chi_{\textrm{RKKY},1}+\chi_{\textrm{RKKY},2}+\chi_{\textrm{RKKY},3}+\chi_{\textrm{RKKY},4} 
\end{pmatrix}+A_{ca} \begin{pmatrix}
	0\\0\\-m_{\textrm{AF},1}+m_{\textrm{AF},2}-m_{\textrm{AF},3}+m_{\textrm{AF},4} 
\end{pmatrix}=\textbf{\textit{H}}_{0} + A_{cc}\textbf{\textit{H}}_{0}\sum_{\textrm{NN}}\chi_{\textrm{RKKY},i}+A_{ca}\sum_{\textrm{NN}}(-1)^{i}m_{\textrm{AF},i}\hat{\textbf{\textit{c}}}$.
This result is calculated for Na nuclei that are above the $ab$-plane, for those which are below the $ab$-plane, $\textbf{\textit{H}}_{\textrm{total}} \approx \textbf{\textit{H}}_{0} + A_{cc}\textbf{\textit{H}}_{0}\sum_{\textrm{NN}}\chi_{\textrm{RKKY},i}-A_{ca}\sum_{\textrm{NN}}(-1)^{i}m_{\textrm{AF},i}\hat{\textbf{\textit{c}}}$. 

%%%%%%%%%%%%%%%%%%%%%%%%%%%%%%%%%%%%%%%%%%%%%%%%%%%%%%%%%%%%%%%%%%%%%%%%%%%%%%%%%%%%%
\subsection{APPENDIX B: Time recovery of spin-lattice and spin-spin relaxation}
\begin{figure}[h]
	\centering
	\includegraphics[scale=0.20]{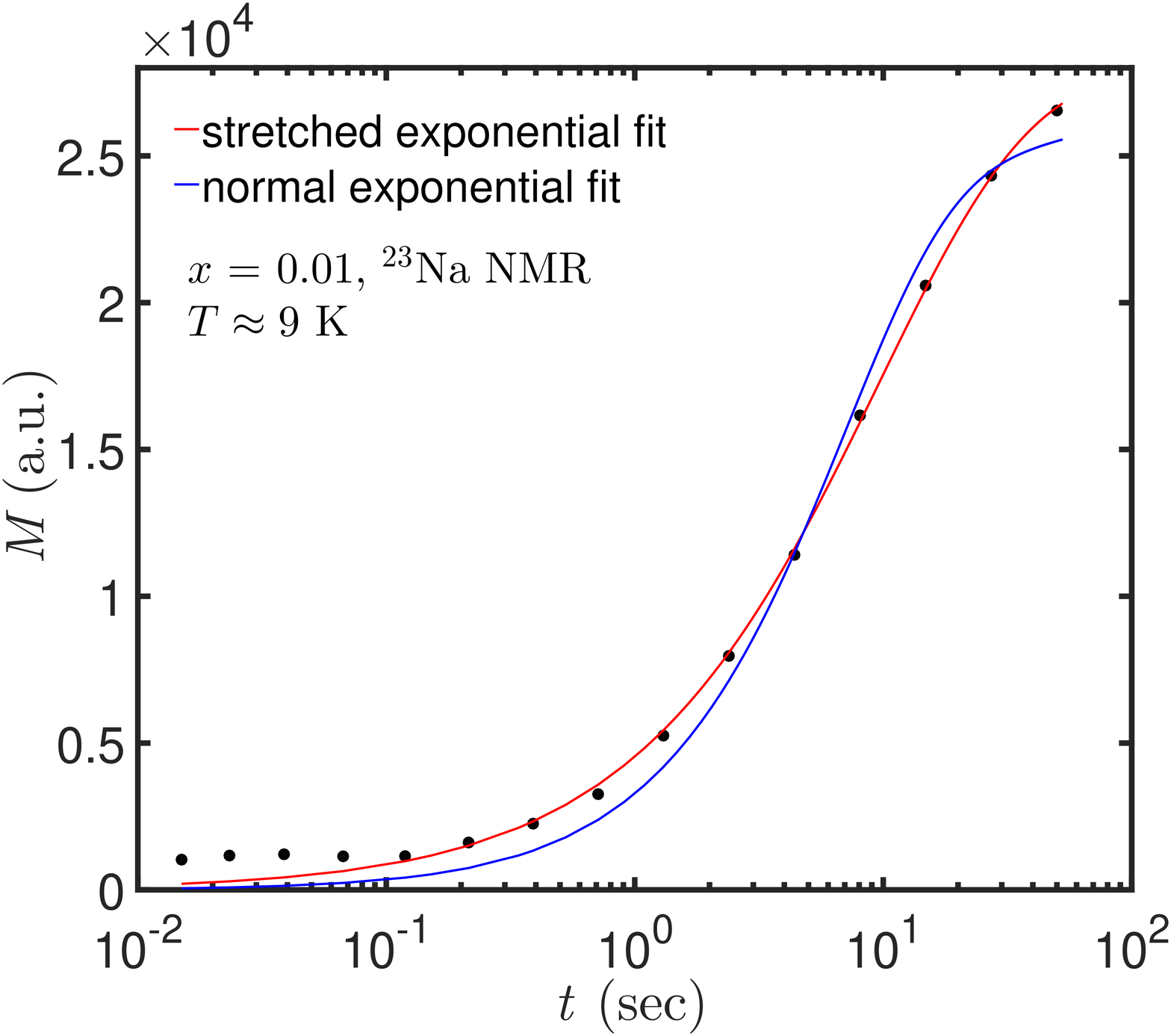}
	\caption{Low-temperature time recovery of spin-lattice relaxation for $x = 0.01$; $H_0 ~||~ c$.  For temperatures approaching $T_\textrm{AF}$, the recovery is better fit to the stretched exponential formula than to the normal recovery.}\label{SFig_StretchedFit}
\end{figure}

For the time recovery of the spin-lattice relaxation, we fit the time dependence of the longitudinal magnetization of the central transition to a stretched exponential formula, 
\begin{equation}\label{SEq_Stretched}
	M(t) = M_0[1-2f(\theta)(0.9\times e^{-(6t/T_1)^{\beta}}+0.1\times e^{-(t/T_1)^{\beta}})]. 
\end{equation}
\noindent where $f(\theta)$ is a function of the tipping angle $\theta$; $f(\theta) = \frac{1-cos(\theta)}{2}$. The exponent $\beta$ arises as a result of a distribution of relaxation rates~\cite{Dio.13,Mit.08,Joh.06}. For $\beta = 1$, Eq.~\ref{SEq_Stretched} is equivalent to the normal recovery formula for the central transition (-1/2 $\leftrightarrow$ 1/2), for $\textit{S} = 3/2$. In Fig.~\ref{SFig_StretchedFit}, a comparison between the normal recovery fit and the stretched exponential fit for $x = 0.01$ is shown. The data clearly fits to the stretched exponential formula better for temperatures close to $T_{\textrm{AF}}$. For $T$ close to and below $T_\textrm{AF}$, $T_{1}$ was measured from non-fully relaxed magnetization. For this approach, we approximate the error margin of the fitted $T_{1}$ to be $\sim$ 4\%, a small error compared to the statistical uncertainty in the fitted value, $\sim$$\pm$20\%.

%comparison between normal and stretched exponential fit

\begin{figure}[h!]
	\centering
	\includegraphics[scale = 0.20]{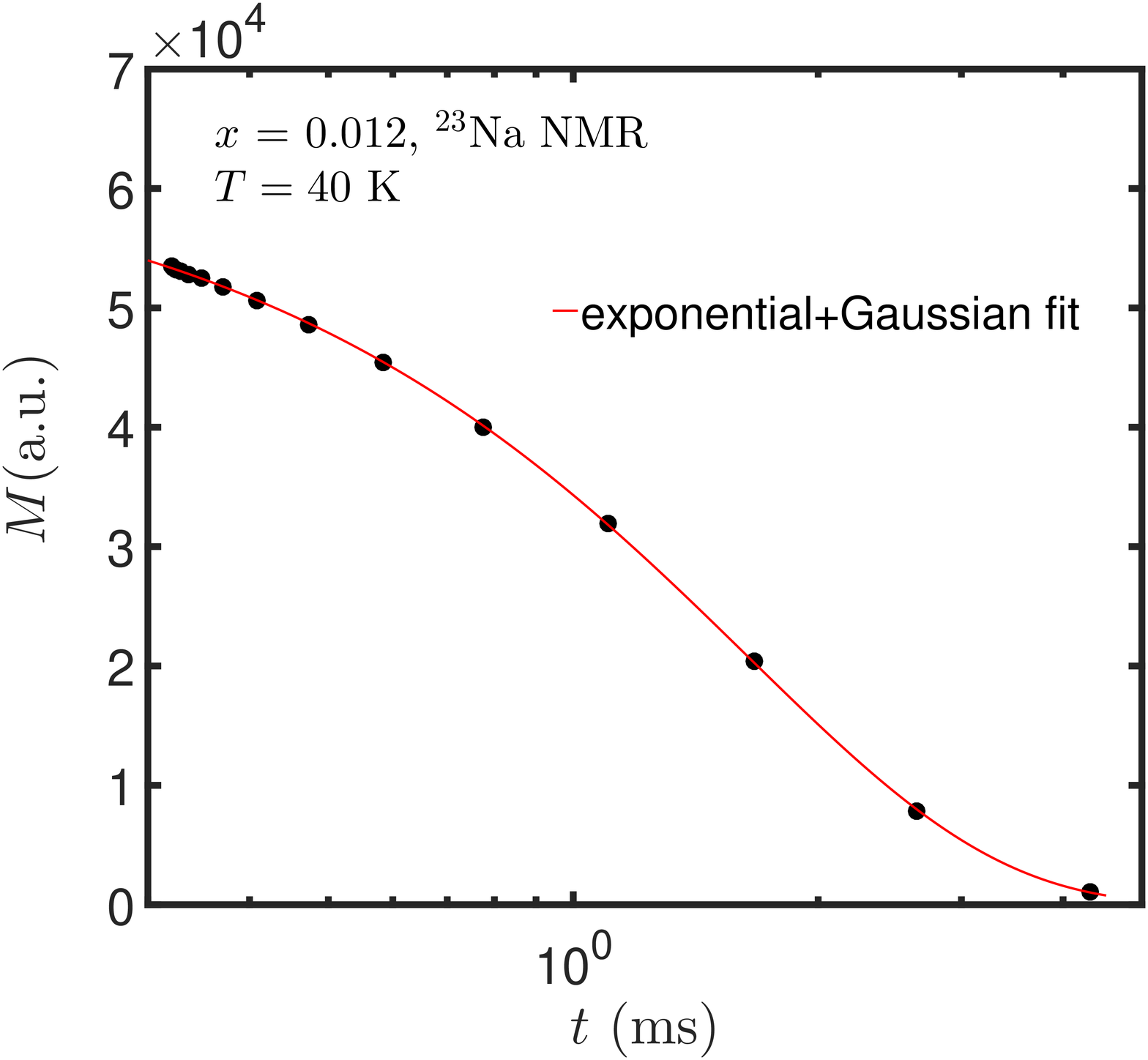}
	\caption{Time recovery of spin-spin relaxation for $x = 0.012$; $H_0 ~||~ c$. The red solid line represents a fit composed of both exponential and Gaussian terms.}\label{SFig_T2Recovery}
\end{figure}

For the time recovery of the spin-spin relaxation, the time dependence of the longitudinal magnetization is fit to 
\begin{equation}\label{SEq_ExpGauss}
	M(t) = M_0\mathrm{exp}(-\frac{t}{T_{2,e}})\mathrm{exp}(-\frac{t^2}{T_{2,g}^2}). 
\end{equation}
\noindent where $T_{2,e}$ and $T_{2,g}$ represent the exponential and Gaussian component of the spin-spin relaxation, respectively. An example of the $T_{2}$ relaxation process is shown in Fig.~\ref{SFig_T2Recovery}.
%time recovery of magnetization due to T2
%%%%%%%%%%%%%%%%%%%%%%%%%%%%%%%%%%%%%%%%%%%%%%%%%%%%%%%%%%%%%%%%%%%%%%%%%%%%%%%%%%
\nolinebreak
\subsection{APPENDIX C: $^{23}$Na Spin-lattice relaxation of heavily Cu-doped compounds}
\begin{figure}[h]
	\centering
	\includegraphics[scale = 0.51]{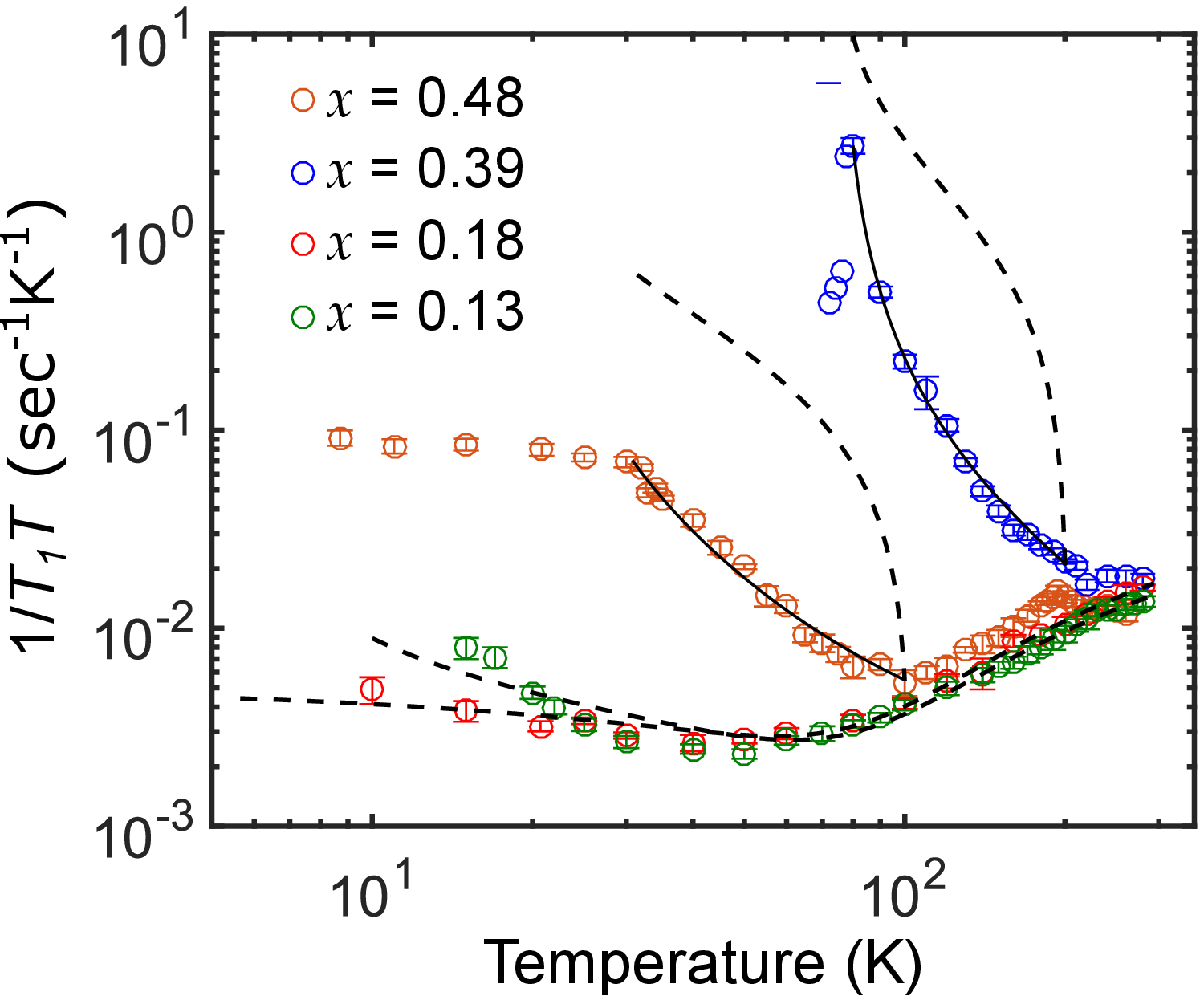}
	\caption{$^{23}$Na spin-lattice relaxation for heavily Cu-doped samples, with $H_{0} = 14~\textrm{T}~ || ~c$. The dashed lines represent fits to a combination of pseudogap and Curie-Weiss terms; $1/T_{1}T = (1/T_1T)_{\textrm{pg}} + (1/T_1T)_{\textrm{AF}} = [A ~\textrm{exp}(-\triangle_{pg}/k_BT) + B] + C / (T - T_{\mathrm{AF}})^{\gamma}$, with $\gamma = 0.5$. The solid lines represent fits to the same terms but with $\gamma$ set as a free parameter.}\label{SFig_OtherSpinRelaxation}
\end{figure}

The $^{23}$Na spin-lattice relaxation ($1/T_{1}T$) for the heavily Cu-doped compounds ($x = 0.13, 0.18,$ 0.39, and 0.48) is shown in Fig.~\ref{SFig_OtherSpinRelaxation}. The dashed lines represent fits to a combination of pseudogap and Curie-Weiss terms,  $1/T_{1}T = (1/T_1T)_{\textrm{pg}} + (1/T_1T)_{\textrm{AF}} = [A ~\textrm{exp}(-\triangle_{pg}/k_BT) + B] + C / (T - T_{\mathrm{AF}})^{0.5}$, as discussed in the main text. The data fit to this formula relatively well for $x = 0.13$ and 0.18; however, $1/T_{1}T$ clearly deviate from this form for $x = 0.39$ and 0.48. Setting $\gamma$ as a free fitting parameter, we obtained $\gamma = 1.5\pm0.1$ and $1.57\pm1.30$ for $x = 0.39$ and 0.48 respectively. This deviation of $\gamma$ from 0.5 indicates that a picture of antiferromagnetic metal is no longer valid for the heavily Cu-doped samples with $x \gtrsim0.3$~\cite{Son.16,Cha.17,Mat.16}. 

%%%%%%%%%%%%%%%%%%%%%%%%%%%%%%%%%%%%%%%%%%%%%%%%%%%%%%%%%%%%%%%%%%%%%%%%%%%%%%%%%%%%%%%
\subsection{APPENDIX D: Temperature-dependent frequency shift of $^{23}$Na satellite resonances}
\begin{figure}[h]
	\centering
	\includegraphics[scale = 0.22]{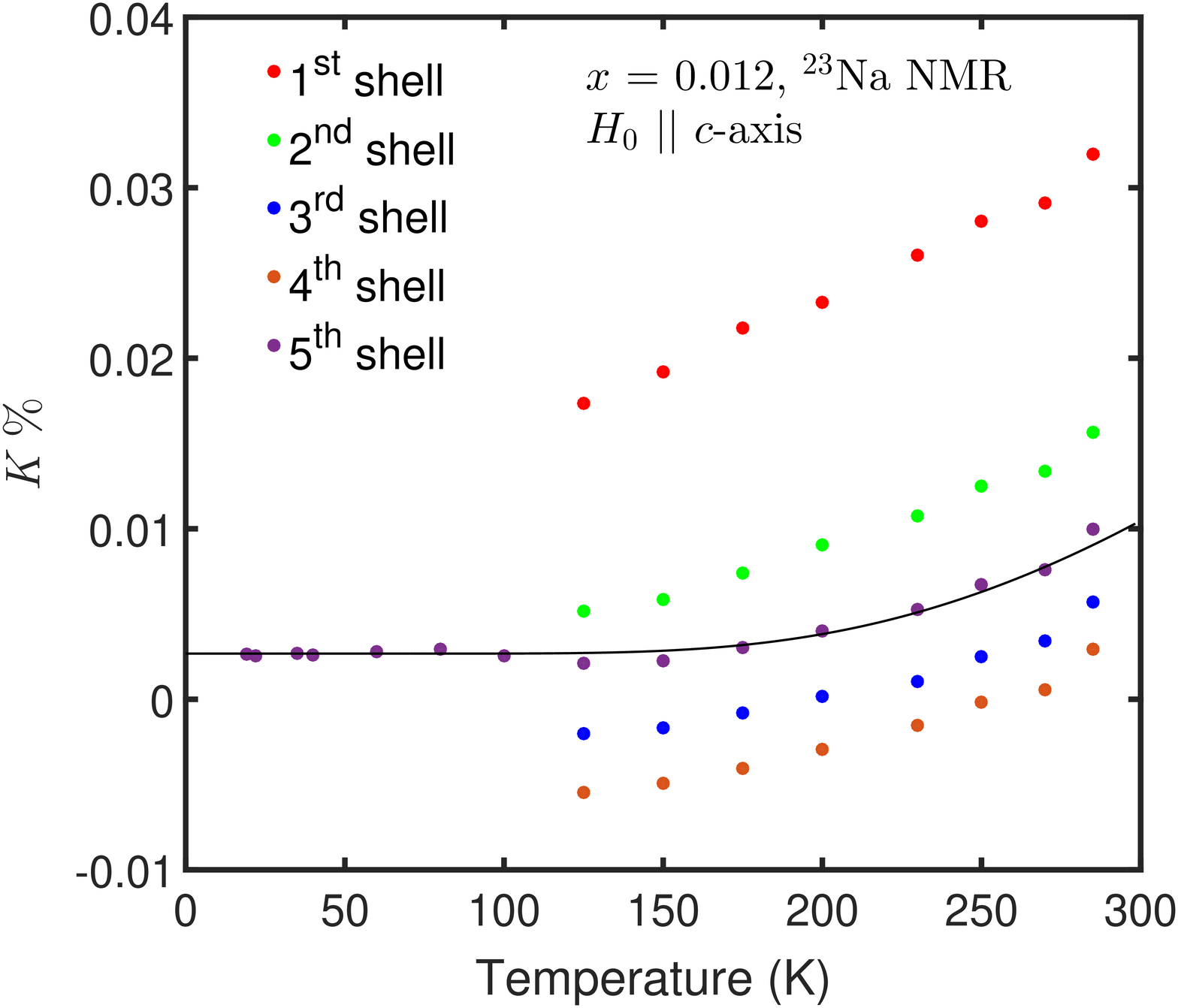}
	\caption{Temperature dependence of $^{23}$Na frequency shift for $x = 0.012$. The data points are color coded to be associated with each satellite, consistent with Fig.~2\,(b), (c), and (e) in the main text. The solid line is a fit to an activation process; $K(T) = A\times \textrm{exp}(-\triangle_{\textrm{pg}}/k_BT)+B$, where $\triangle_{\textrm{pg}}$ is the spin pseudogap, $k_{B}$ is the Boltzmann constant, and $A$ and $B$ are constants.}\label{SFig_Freqshift}
\end{figure}

\begin{figure}[h]
	\centering
	\includegraphics[scale = 0.20]{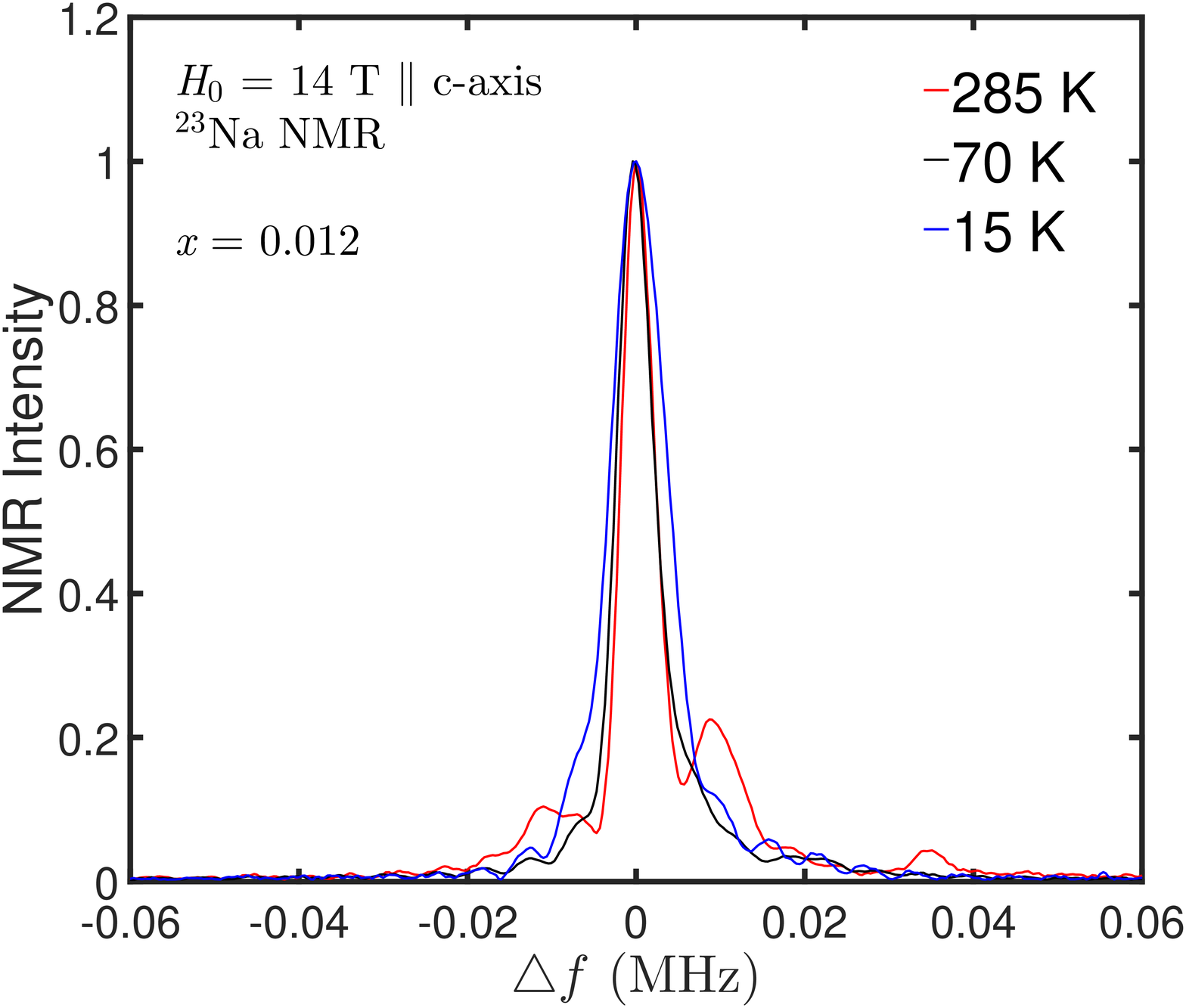}
	\caption{Line-broadening of $^{23}$Na spectrum. All satellites broaden significantly as the temperature approaches $T_\textrm{AF}$, rendering them difficult to be resolved at low temperatures. }\label{SFig_Linebroadening}
\end{figure}

The frequency shift, $K$, is defined as the percentage shift relative to the Larmor frequency ($f_{L}$);  $K = \frac{f-f_{L}}{\omega_{L}}*100\%$. For $x = 0.012$, the temperature dependence of the frequency shift for different $^{23}$Na satellite resonances and the main spectral component is shown in Fig.~\ref{SFig_Freqshift}. Each satellite component is contributed by a group of Na nuclei at a certain distance from a Cu dopant, as discussed in the main text. For temperatures down to $\sim125~\textrm{K}$, the frequency shift of each satellite follows almost the same temperature dependence, indicating that the RKKY interaction is largely temperature-independent at high temperatures. For $T<125~\textrm{K}$, line broadening renders the satellites difficult to resolve, as shown by a comparison of the room-temperature $^{23}$Na spectrum with those taken at lower temperatures; Fig~\ref{SFig_Linebroadening}. The main peak is largely immune from the line broadening and can be clearly resolved even in the AFM phase. Consequently, we show the frequency shift data for this spectral component down to $\sim$20~K in Fig.~\ref{SFig_Freqshift}.

%%%%%%%%%%%%%%%%%%%%%%%%%%%%% End of Appendix %%%%%%%%%%%%%%%%%%%%%%%%%%%%%%%%%%%%%%

\bibliography{maintext}

\begin{thebibliography}{56}%
\makeatletter
\providecommand \@ifxundefined [1]{%
 \@ifx{#1\undefined}
}%
\providecommand \@ifnum [1]{%
 \ifnum #1\expandafter \@firstoftwo
 \else \expandafter \@secondoftwo
 \fi
}%
\providecommand \@ifx [1]{%
 \ifx #1\expandafter \@firstoftwo
 \else \expandafter \@secondoftwo
 \fi
}%
\providecommand \natexlab [1]{#1}%
\providecommand \enquote  [1]{``#1''}%
\providecommand \bibnamefont  [1]{#1}%
\providecommand \bibfnamefont [1]{#1}%
\providecommand \citenamefont [1]{#1}%
\providecommand \href@noop [0]{\@secondoftwo}%
\providecommand \href [0]{\begingroup \@sanitize@url \@href}%
\providecommand \@href[1]{\@@startlink{#1}\@@href}%
\providecommand \@@href[1]{\endgroup#1\@@endlink}%
\providecommand \@sanitize@url [0]{\catcode `\\12\catcode `\$12\catcode
  `\&12\catcode `\#12\catcode `\^12\catcode `\_12\catcode `\%12\relax}%
\providecommand \@@startlink[1]{}%
\providecommand \@@endlink[0]{}%
\providecommand \url  [0]{\begingroup\@sanitize@url \@url }%
\providecommand \@url [1]{\endgroup\@href {#1}{\urlprefix }}%
\providecommand \urlprefix  [0]{URL }%
\providecommand \Eprint [0]{\href }%
\providecommand \doibase [0]{https://doi.org/}%
\providecommand \selectlanguage [0]{\@gobble}%
\providecommand \bibinfo  [0]{\@secondoftwo}%
\providecommand \bibfield  [0]{\@secondoftwo}%
\providecommand \translation [1]{[#1]}%
\providecommand \BibitemOpen [0]{}%
\providecommand \bibitemStop [0]{}%
\providecommand \bibitemNoStop [0]{.\EOS\space}%
\providecommand \EOS [0]{\spacefactor3000\relax}%
\providecommand \BibitemShut  [1]{\csname bibitem#1\endcsname}%
\let\auto@bib@innerbib\@empty
%</preamble>
\bibitem [{\citenamefont {Boyce}\ and\ \citenamefont
  {Slichter}(1974)}]{Boy.74}%
  \BibitemOpen
  \bibfield  {author} {\bibinfo {author} {\bibfnamefont {J.~B.}\ \bibnamefont
  {Boyce}}\ and\ \bibinfo {author} {\bibfnamefont {C.~P.}\ \bibnamefont
  {Slichter}},\ }\href {https://doi.org/10.1103/PhysRevLett.32.61} {\bibfield
  {journal} {\bibinfo  {journal} {Phys. Rev. Lett.}\ }\textbf {\bibinfo
  {volume} {32}},\ \bibinfo {pages} {61} (\bibinfo {year} {1974})}\BibitemShut
  {NoStop}%
\bibitem [{\citenamefont {Alloul}(1974)}]{All.74}%
  \BibitemOpen
  \bibfield  {author} {\bibinfo {author} {\bibfnamefont {H.}~\bibnamefont
  {Alloul}},\ }\href {https://doi.org/10.1088/0305-4608/4/9/021} {\bibfield
  {journal} {\bibinfo  {journal} {Journal of Physics F: Metal Physics}\
  }\textbf {\bibinfo {volume} {4}},\ \bibinfo {pages} {1501} (\bibinfo {year}
  {1974})}\BibitemShut {NoStop}%
\bibitem [{\citenamefont {Alloul}(1977)}]{All.77}%
  \BibitemOpen
  \bibfield  {author} {\bibinfo {author} {\bibfnamefont {H.}~\bibnamefont
  {Alloul}},\ }\href
  {https://doi.org/https://doi.org/10.1016/0378-4363(77)90382-5} {\bibfield
  {journal} {\bibinfo  {journal} {Physica B+C}\ }\textbf {\bibinfo {volume}
  {86-88}},\ \bibinfo {pages} {449} (\bibinfo {year} {1977})}\BibitemShut
  {NoStop}%
\bibitem [{\citenamefont {Paschen}\ and\ \citenamefont {Si}(2021)}]{Pas.21}%
  \BibitemOpen
  \bibfield  {author} {\bibinfo {author} {\bibfnamefont {S.}~\bibnamefont
  {Paschen}}\ and\ \bibinfo {author} {\bibfnamefont {Q.}~\bibnamefont {Si}},\
  }\href@noop {} {\bibfield  {journal} {\bibinfo  {journal} {Nature Reviews
  Physics}\ }\textbf {\bibinfo {volume} {3}},\ \bibinfo {pages} {9} (\bibinfo
  {year} {2021})}\BibitemShut {NoStop}%
\bibitem [{\citenamefont {Varma}(1976)}]{Var.76}%
  \BibitemOpen
  \bibfield  {author} {\bibinfo {author} {\bibfnamefont {C.~M.}\ \bibnamefont
  {Varma}},\ }\href {https://doi.org/10.1103/RevModPhys.48.219} {\bibfield
  {journal} {\bibinfo  {journal} {Rev. Mod. Phys.}\ }\textbf {\bibinfo {volume}
  {48}},\ \bibinfo {pages} {219} (\bibinfo {year} {1976})}\BibitemShut
  {NoStop}%
\bibitem [{\citenamefont {Gegenwart}\ \emph {et~al.}(2008)\citenamefont
  {Gegenwart}, \citenamefont {Si},\ and\ \citenamefont {Steglich}}]{Geg.08}%
  \BibitemOpen
  \bibfield  {author} {\bibinfo {author} {\bibfnamefont {P.}~\bibnamefont
  {Gegenwart}}, \bibinfo {author} {\bibfnamefont {Q.}~\bibnamefont {Si}},\ and\
  \bibinfo {author} {\bibfnamefont {F.}~\bibnamefont {Steglich}},\ }\href@noop
  {} {\bibfield  {journal} {\bibinfo  {journal} {Nature Phys}\ }\textbf
  {\bibinfo {volume} {4}},\ \bibinfo {pages} {186} (\bibinfo {year}
  {2008})}\BibitemShut {NoStop}%
\bibitem [{\citenamefont {Hirayama}\ \emph {et~al.}(2015)\citenamefont
  {Hirayama}, \citenamefont {Misawa}, \citenamefont {Miyake},\ and\
  \citenamefont {Imada}}]{Hir.15}%
  \BibitemOpen
  \bibfield  {author} {\bibinfo {author} {\bibfnamefont {M.}~\bibnamefont
  {Hirayama}}, \bibinfo {author} {\bibfnamefont {T.}~\bibnamefont {Misawa}},
  \bibinfo {author} {\bibfnamefont {T.}~\bibnamefont {Miyake}},\ and\ \bibinfo
  {author} {\bibfnamefont {M.}~\bibnamefont {Imada}},\ }\href
  {https://doi.org/10.7566/JPSJ.84.093703} {\bibfield  {journal} {\bibinfo
  {journal} {Journal of the Physical Society of Japan}\ }\textbf {\bibinfo
  {volume} {84}},\ \bibinfo {pages} {093703} (\bibinfo {year} {2015})},\
  \Eprint {https://arxiv.org/abs/https://doi.org/10.7566/JPSJ.84.093703}
  {https://doi.org/10.7566/JPSJ.84.093703} \BibitemShut {NoStop}%
\bibitem [{\citenamefont {Akbari}\ \emph {et~al.}(2011)\citenamefont {Akbari},
  \citenamefont {Eremin},\ and\ \citenamefont {Thalmeier}}]{Akb.11}%
  \BibitemOpen
  \bibfield  {author} {\bibinfo {author} {\bibfnamefont {A.}~\bibnamefont
  {Akbari}}, \bibinfo {author} {\bibfnamefont {I.}~\bibnamefont {Eremin}},\
  and\ \bibinfo {author} {\bibfnamefont {P.}~\bibnamefont {Thalmeier}},\ }\href
  {https://doi.org/10.1103/PhysRevB.84.134513} {\bibfield  {journal} {\bibinfo
  {journal} {Phys. Rev. B}\ }\textbf {\bibinfo {volume} {84}},\ \bibinfo
  {pages} {134513} (\bibinfo {year} {2011})}\BibitemShut {NoStop}%
\bibitem [{\citenamefont {Zapf}\ \emph {et~al.}(2011)\citenamefont {Zapf},
  \citenamefont {Wu}, \citenamefont {Bogani}, \citenamefont {Jeevan},
  \citenamefont {Gegenwart},\ and\ \citenamefont {Dressel}}]{Zap.11}%
  \BibitemOpen
  \bibfield  {author} {\bibinfo {author} {\bibfnamefont {S.}~\bibnamefont
  {Zapf}}, \bibinfo {author} {\bibfnamefont {D.}~\bibnamefont {Wu}}, \bibinfo
  {author} {\bibfnamefont {L.}~\bibnamefont {Bogani}}, \bibinfo {author}
  {\bibfnamefont {H.~S.}\ \bibnamefont {Jeevan}}, \bibinfo {author}
  {\bibfnamefont {P.}~\bibnamefont {Gegenwart}},\ and\ \bibinfo {author}
  {\bibfnamefont {M.}~\bibnamefont {Dressel}},\ }\href
  {https://doi.org/10.1103/PhysRevB.84.140503} {\bibfield  {journal} {\bibinfo
  {journal} {Phys. Rev. B}\ }\textbf {\bibinfo {volume} {84}},\ \bibinfo
  {pages} {140503} (\bibinfo {year} {2011})}\BibitemShut {NoStop}%
\bibitem [{\citenamefont {Moroni}\ \emph {et~al.}(2017)\citenamefont {Moroni},
  \citenamefont {Carretta}, \citenamefont {Allodi}, \citenamefont {De~Renzi},
  \citenamefont {Gastiasoro}, \citenamefont {Andersen}, \citenamefont
  {Materne}, \citenamefont {Klauss}, \citenamefont {Kobayashi}, \citenamefont
  {Sato},\ and\ \citenamefont {Sanna}}]{Mor.17}%
  \BibitemOpen
  \bibfield  {author} {\bibinfo {author} {\bibfnamefont {M.}~\bibnamefont
  {Moroni}}, \bibinfo {author} {\bibfnamefont {P.}~\bibnamefont {Carretta}},
  \bibinfo {author} {\bibfnamefont {G.}~\bibnamefont {Allodi}}, \bibinfo
  {author} {\bibfnamefont {R.}~\bibnamefont {De~Renzi}}, \bibinfo {author}
  {\bibfnamefont {M.~N.}\ \bibnamefont {Gastiasoro}}, \bibinfo {author}
  {\bibfnamefont {B.~M.}\ \bibnamefont {Andersen}}, \bibinfo {author}
  {\bibfnamefont {P.}~\bibnamefont {Materne}}, \bibinfo {author} {\bibfnamefont
  {H.-H.}\ \bibnamefont {Klauss}}, \bibinfo {author} {\bibfnamefont
  {Y.}~\bibnamefont {Kobayashi}}, \bibinfo {author} {\bibfnamefont
  {M.}~\bibnamefont {Sato}},\ and\ \bibinfo {author} {\bibfnamefont
  {S.}~\bibnamefont {Sanna}},\ }\href
  {https://doi.org/10.1103/PhysRevB.95.180501} {\bibfield  {journal} {\bibinfo
  {journal} {Phys. Rev. B}\ }\textbf {\bibinfo {volume} {95}},\ \bibinfo
  {pages} {180501} (\bibinfo {year} {2017})}\BibitemShut {NoStop}%
\bibitem [{\citenamefont {LeBoeuf}\ \emph {et~al.}(2014)\citenamefont
  {LeBoeuf}, \citenamefont {Texier}, \citenamefont {Boselli}, \citenamefont
  {Forget}, \citenamefont {Colson},\ and\ \citenamefont {Bobroff}}]{Leb.14}%
  \BibitemOpen
  \bibfield  {author} {\bibinfo {author} {\bibfnamefont {D.}~\bibnamefont
  {LeBoeuf}}, \bibinfo {author} {\bibfnamefont {Y.}~\bibnamefont {Texier}},
  \bibinfo {author} {\bibfnamefont {M.}~\bibnamefont {Boselli}}, \bibinfo
  {author} {\bibfnamefont {A.}~\bibnamefont {Forget}}, \bibinfo {author}
  {\bibfnamefont {D.}~\bibnamefont {Colson}},\ and\ \bibinfo {author}
  {\bibfnamefont {J.}~\bibnamefont {Bobroff}},\ }\href
  {https://doi.org/10.1103/PhysRevB.89.035114} {\bibfield  {journal} {\bibinfo
  {journal} {Phys. Rev. B}\ }\textbf {\bibinfo {volume} {89}},\ \bibinfo
  {pages} {035114} (\bibinfo {year} {2014})}\BibitemShut {NoStop}%
\bibitem [{\citenamefont {Wang}\ \emph {et~al.}(2013)\citenamefont {Wang},
  \citenamefont {Lin}, \citenamefont {Cheng}, \citenamefont {Ye}, \citenamefont
  {Chen}, \citenamefont {Ma}, \citenamefont {Lu}, \citenamefont {Lei},
  \citenamefont {Luo},\ and\ \citenamefont {Chen}}]{Wan.13}%
  \BibitemOpen
  \bibfield  {author} {\bibinfo {author} {\bibfnamefont {A.~F.}\ \bibnamefont
  {Wang}}, \bibinfo {author} {\bibfnamefont {J.~J.}\ \bibnamefont {Lin}},
  \bibinfo {author} {\bibfnamefont {P.}~\bibnamefont {Cheng}}, \bibinfo
  {author} {\bibfnamefont {G.~J.}\ \bibnamefont {Ye}}, \bibinfo {author}
  {\bibfnamefont {F.}~\bibnamefont {Chen}}, \bibinfo {author} {\bibfnamefont
  {J.~Q.}\ \bibnamefont {Ma}}, \bibinfo {author} {\bibfnamefont {X.~F.}\
  \bibnamefont {Lu}}, \bibinfo {author} {\bibfnamefont {B.}~\bibnamefont
  {Lei}}, \bibinfo {author} {\bibfnamefont {X.~G.}\ \bibnamefont {Luo}},\ and\
  \bibinfo {author} {\bibfnamefont {X.~H.}\ \bibnamefont {Chen}},\ }\href
  {https://doi.org/10.1103/PhysRevB.88.094516} {\bibfield  {journal} {\bibinfo
  {journal} {Phys. Rev. B}\ }\textbf {\bibinfo {volume} {88}},\ \bibinfo
  {pages} {094516} (\bibinfo {year} {2013})}\BibitemShut {NoStop}%
\bibitem [{\citenamefont {Ma}\ \emph {et~al.}(2011)\citenamefont {Ma},
  \citenamefont {Chen}, \citenamefont {Yao}, \citenamefont {Zhang},
  \citenamefont {Zhang}, \citenamefont {Xia},\ and\ \citenamefont
  {Yu}}]{Ma.11}%
  \BibitemOpen
  \bibfield  {author} {\bibinfo {author} {\bibfnamefont {L.}~\bibnamefont
  {Ma}}, \bibinfo {author} {\bibfnamefont {G.~F.}\ \bibnamefont {Chen}},
  \bibinfo {author} {\bibfnamefont {D.-X.}\ \bibnamefont {Yao}}, \bibinfo
  {author} {\bibfnamefont {J.}~\bibnamefont {Zhang}}, \bibinfo {author}
  {\bibfnamefont {S.}~\bibnamefont {Zhang}}, \bibinfo {author} {\bibfnamefont
  {T.-L.}\ \bibnamefont {Xia}},\ and\ \bibinfo {author} {\bibfnamefont
  {W.}~\bibnamefont {Yu}},\ }\href {https://doi.org/10.1103/PhysRevB.83.132501}
  {\bibfield  {journal} {\bibinfo  {journal} {Phys. Rev. B}\ }\textbf {\bibinfo
  {volume} {83}},\ \bibinfo {pages} {132501} (\bibinfo {year}
  {2011})}\BibitemShut {NoStop}%
\bibitem [{\citenamefont {Song}\ \emph {et~al.}(2016)\citenamefont {Song},
  \citenamefont {Yamani}, \citenamefont {Cao}, \citenamefont {Li},
  \citenamefont {Zhang}, \citenamefont {Chen}, \citenamefont {Huang},
  \citenamefont {Wu}, \citenamefont {Tao}, \citenamefont {Zhu} \emph
  {et~al.}}]{Son.16}%
  \BibitemOpen
  \bibfield  {author} {\bibinfo {author} {\bibfnamefont {Y.}~\bibnamefont
  {Song}}, \bibinfo {author} {\bibfnamefont {Z.}~\bibnamefont {Yamani}},
  \bibinfo {author} {\bibfnamefont {C.}~\bibnamefont {Cao}}, \bibinfo {author}
  {\bibfnamefont {Y.}~\bibnamefont {Li}}, \bibinfo {author} {\bibfnamefont
  {C.}~\bibnamefont {Zhang}}, \bibinfo {author} {\bibfnamefont {J.~S.}\
  \bibnamefont {Chen}}, \bibinfo {author} {\bibfnamefont {Q.}~\bibnamefont
  {Huang}}, \bibinfo {author} {\bibfnamefont {H.}~\bibnamefont {Wu}}, \bibinfo
  {author} {\bibfnamefont {J.}~\bibnamefont {Tao}}, \bibinfo {author}
  {\bibfnamefont {Y.}~\bibnamefont {Zhu}}, \emph {et~al.},\ }\href@noop {}
  {\bibfield  {journal} {\bibinfo  {journal} {Nature communications}\ }\textbf
  {\bibinfo {volume} {7}},\ \bibinfo {pages} {13879} (\bibinfo {year}
  {2016})}\BibitemShut {NoStop}%
\bibitem [{\citenamefont {Matt}\ \emph {et~al.}(2016)\citenamefont {Matt},
  \citenamefont {Xu}, \citenamefont {Lv}, \citenamefont {Ma}, \citenamefont
  {Bisti}, \citenamefont {Park}, \citenamefont {Shang}, \citenamefont {Cao},
  \citenamefont {Song}, \citenamefont {Nevidomskyy}, \citenamefont {Dai},
  \citenamefont {Patthey}, \citenamefont {Plumb}, \citenamefont {Radovic},
  \citenamefont {Mesot},\ and\ \citenamefont {Shi}}]{Mat.16}%
  \BibitemOpen
  \bibfield  {author} {\bibinfo {author} {\bibfnamefont {C.~E.}\ \bibnamefont
  {Matt}}, \bibinfo {author} {\bibfnamefont {N.}~\bibnamefont {Xu}}, \bibinfo
  {author} {\bibfnamefont {B.}~\bibnamefont {Lv}}, \bibinfo {author}
  {\bibfnamefont {J.}~\bibnamefont {Ma}}, \bibinfo {author} {\bibfnamefont
  {F.}~\bibnamefont {Bisti}}, \bibinfo {author} {\bibfnamefont
  {J.}~\bibnamefont {Park}}, \bibinfo {author} {\bibfnamefont {T.}~\bibnamefont
  {Shang}}, \bibinfo {author} {\bibfnamefont {C.}~\bibnamefont {Cao}}, \bibinfo
  {author} {\bibfnamefont {Y.}~\bibnamefont {Song}}, \bibinfo {author}
  {\bibfnamefont {A.~H.}\ \bibnamefont {Nevidomskyy}}, \bibinfo {author}
  {\bibfnamefont {P.}~\bibnamefont {Dai}}, \bibinfo {author} {\bibfnamefont
  {L.}~\bibnamefont {Patthey}}, \bibinfo {author} {\bibfnamefont {N.~C.}\
  \bibnamefont {Plumb}}, \bibinfo {author} {\bibfnamefont {M.}~\bibnamefont
  {Radovic}}, \bibinfo {author} {\bibfnamefont {J.}~\bibnamefont {Mesot}},\
  and\ \bibinfo {author} {\bibfnamefont {M.}~\bibnamefont {Shi}},\ }\href
  {https://doi.org/10.1103/PhysRevLett.117.097001} {\bibfield  {journal}
  {\bibinfo  {journal} {Phys. Rev. Lett.}\ }\textbf {\bibinfo {volume} {117}},\
  \bibinfo {pages} {097001} (\bibinfo {year} {2016})}\BibitemShut {NoStop}%
\bibitem [{\citenamefont {Zhang}\ \emph {et~al.}(2017)\citenamefont {Zhang},
  \citenamefont {He}, \citenamefont {Mei}, \citenamefont {Liu},\ and\
  \citenamefont {Liu}}]{Zha.17}%
  \BibitemOpen
  \bibfield  {author} {\bibinfo {author} {\bibfnamefont {S.}~\bibnamefont
  {Zhang}}, \bibinfo {author} {\bibfnamefont {Y.}~\bibnamefont {He}}, \bibinfo
  {author} {\bibfnamefont {J.-W.}\ \bibnamefont {Mei}}, \bibinfo {author}
  {\bibfnamefont {F.}~\bibnamefont {Liu}},\ and\ \bibinfo {author}
  {\bibfnamefont {Z.}~\bibnamefont {Liu}},\ }\href
  {https://doi.org/10.1103/PhysRevB.96.245128} {\bibfield  {journal} {\bibinfo
  {journal} {Phys. Rev. B}\ }\textbf {\bibinfo {volume} {96}},\ \bibinfo
  {pages} {245128} (\bibinfo {year} {2017})}\BibitemShut {NoStop}%
\bibitem [{\citenamefont {Xin}\ \emph {et~al.}(2019)\citenamefont {Xin},
  \citenamefont {Stolt}, \citenamefont {Lee}, \citenamefont {Song},
  \citenamefont {Dai},\ and\ \citenamefont {Halperin}}]{Xin.19}%
  \BibitemOpen
  \bibfield  {author} {\bibinfo {author} {\bibfnamefont {Y.}~\bibnamefont
  {Xin}}, \bibinfo {author} {\bibfnamefont {I.}~\bibnamefont {Stolt}}, \bibinfo
  {author} {\bibfnamefont {J.~A.}\ \bibnamefont {Lee}}, \bibinfo {author}
  {\bibfnamefont {Y.}~\bibnamefont {Song}}, \bibinfo {author} {\bibfnamefont
  {P.}~\bibnamefont {Dai}},\ and\ \bibinfo {author} {\bibfnamefont {W.~P.}\
  \bibnamefont {Halperin}},\ }\href
  {https://doi.org/10.1103/PhysRevB.99.155114} {\bibfield  {journal} {\bibinfo
  {journal} {Phys. Rev. B}\ }\textbf {\bibinfo {volume} {99}},\ \bibinfo
  {pages} {155114} (\bibinfo {year} {2019})}\BibitemShut {NoStop}%
\bibitem [{\citenamefont {Xin}\ \emph {et~al.}(2020)\citenamefont {Xin},
  \citenamefont {Stolt}, \citenamefont {Song}, \citenamefont {Dai},\ and\
  \citenamefont {Halperin}}]{Xin.20}%
  \BibitemOpen
  \bibfield  {author} {\bibinfo {author} {\bibfnamefont {Y.}~\bibnamefont
  {Xin}}, \bibinfo {author} {\bibfnamefont {I.}~\bibnamefont {Stolt}}, \bibinfo
  {author} {\bibfnamefont {Y.}~\bibnamefont {Song}}, \bibinfo {author}
  {\bibfnamefont {P.}~\bibnamefont {Dai}},\ and\ \bibinfo {author}
  {\bibfnamefont {W.~P.}\ \bibnamefont {Halperin}},\ }\href
  {https://doi.org/10.1103/PhysRevB.101.064410} {\bibfield  {journal} {\bibinfo
   {journal} {Phys. Rev. B}\ }\textbf {\bibinfo {volume} {101}},\ \bibinfo
  {pages} {064410} (\bibinfo {year} {2020})}\BibitemShut {NoStop}%
\bibitem [{\citenamefont {Song}\ \emph {et~al.}(2021)\citenamefont {Song},
  \citenamefont {Wang}, \citenamefont {Paris}, \citenamefont {Lu},
  \citenamefont {Pelliciari}, \citenamefont {Tseng}, \citenamefont {Huang},
  \citenamefont {McNally}, \citenamefont {Dantz}, \citenamefont {Cao},
  \citenamefont {Yu}, \citenamefont {Birgeneau}, \citenamefont {Schmitt},\ and\
  \citenamefont {Dai}}]{Son.21}%
  \BibitemOpen
  \bibfield  {author} {\bibinfo {author} {\bibfnamefont {Y.}~\bibnamefont
  {Song}}, \bibinfo {author} {\bibfnamefont {W.}~\bibnamefont {Wang}}, \bibinfo
  {author} {\bibfnamefont {E.}~\bibnamefont {Paris}}, \bibinfo {author}
  {\bibfnamefont {X.}~\bibnamefont {Lu}}, \bibinfo {author} {\bibfnamefont
  {J.}~\bibnamefont {Pelliciari}}, \bibinfo {author} {\bibfnamefont
  {Y.}~\bibnamefont {Tseng}}, \bibinfo {author} {\bibfnamefont
  {Y.}~\bibnamefont {Huang}}, \bibinfo {author} {\bibfnamefont
  {D.}~\bibnamefont {McNally}}, \bibinfo {author} {\bibfnamefont
  {M.}~\bibnamefont {Dantz}}, \bibinfo {author} {\bibfnamefont
  {C.}~\bibnamefont {Cao}}, \bibinfo {author} {\bibfnamefont {R.}~\bibnamefont
  {Yu}}, \bibinfo {author} {\bibfnamefont {R.~J.}\ \bibnamefont {Birgeneau}},
  \bibinfo {author} {\bibfnamefont {T.}~\bibnamefont {Schmitt}},\ and\ \bibinfo
  {author} {\bibfnamefont {P.}~\bibnamefont {Dai}},\ }\href
  {https://doi.org/10.1103/PhysRevB.103.075112} {\bibfield  {journal} {\bibinfo
   {journal} {Phys. Rev. B}\ }\textbf {\bibinfo {volume} {103}},\ \bibinfo
  {pages} {075112} (\bibinfo {year} {2021})}\BibitemShut {NoStop}%
\bibitem [{\citenamefont {Cui}\ \emph {et~al.}(2013)\citenamefont {Cui},
  \citenamefont {Kong}, \citenamefont {Ju}, \citenamefont {Wu}, \citenamefont
  {Wang}, \citenamefont {Luo}, \citenamefont {Chen}, \citenamefont {Zhang},\
  and\ \citenamefont {Sun}}]{Cui.13}%
  \BibitemOpen
  \bibfield  {author} {\bibinfo {author} {\bibfnamefont {S.~T.}\ \bibnamefont
  {Cui}}, \bibinfo {author} {\bibfnamefont {S.}~\bibnamefont {Kong}}, \bibinfo
  {author} {\bibfnamefont {S.~L.}\ \bibnamefont {Ju}}, \bibinfo {author}
  {\bibfnamefont {P.}~\bibnamefont {Wu}}, \bibinfo {author} {\bibfnamefont
  {A.~F.}\ \bibnamefont {Wang}}, \bibinfo {author} {\bibfnamefont {X.~G.}\
  \bibnamefont {Luo}}, \bibinfo {author} {\bibfnamefont {X.~H.}\ \bibnamefont
  {Chen}}, \bibinfo {author} {\bibfnamefont {G.~B.}\ \bibnamefont {Zhang}},\
  and\ \bibinfo {author} {\bibfnamefont {Z.}~\bibnamefont {Sun}},\ }\href
  {https://doi.org/10.1103/PhysRevB.88.245112} {\bibfield  {journal} {\bibinfo
  {journal} {Phys. Rev. B}\ }\textbf {\bibinfo {volume} {88}},\ \bibinfo
  {pages} {245112} (\bibinfo {year} {2013})}\BibitemShut {NoStop}%
\bibitem [{\citenamefont {Yi}\ \emph {et~al.}(2012)\citenamefont {Yi},
  \citenamefont {Lu}, \citenamefont {Moore}, \citenamefont {Kihou},
  \citenamefont {Lee}, \citenamefont {Iyo}, \citenamefont {Eisaki},
  \citenamefont {Yoshida}, \citenamefont {Fujimori},\ and\ \citenamefont
  {Shen}}]{Yi.12}%
  \BibitemOpen
  \bibfield  {author} {\bibinfo {author} {\bibfnamefont {M.}~\bibnamefont
  {Yi}}, \bibinfo {author} {\bibfnamefont {D.~H.}\ \bibnamefont {Lu}}, \bibinfo
  {author} {\bibfnamefont {R.~G.}\ \bibnamefont {Moore}}, \bibinfo {author}
  {\bibfnamefont {K.}~\bibnamefont {Kihou}}, \bibinfo {author} {\bibfnamefont
  {C.-H.}\ \bibnamefont {Lee}}, \bibinfo {author} {\bibfnamefont
  {A.}~\bibnamefont {Iyo}}, \bibinfo {author} {\bibfnamefont {H.}~\bibnamefont
  {Eisaki}}, \bibinfo {author} {\bibfnamefont {T.}~\bibnamefont {Yoshida}},
  \bibinfo {author} {\bibfnamefont {A.}~\bibnamefont {Fujimori}},\ and\
  \bibinfo {author} {\bibfnamefont {Z.-X.}\ \bibnamefont {Shen}},\ }\href
  {https://doi.org/10.1088/1367-2630/14/7/073019} {\bibfield  {journal}
  {\bibinfo  {journal} {New Journal of Physics}\ }\textbf {\bibinfo {volume}
  {14}},\ \bibinfo {pages} {073019} (\bibinfo {year} {2012})}\BibitemShut
  {NoStop}%
\bibitem [{\citenamefont {Zhang}\ \emph {et~al.}(2010)\citenamefont {Zhang},
  \citenamefont {Opahle}, \citenamefont {Jeschke},\ and\ \citenamefont
  {Valent\'{\i}}}]{Zha.10}%
  \BibitemOpen
  \bibfield  {author} {\bibinfo {author} {\bibfnamefont {Y.-Z.}\ \bibnamefont
  {Zhang}}, \bibinfo {author} {\bibfnamefont {I.}~\bibnamefont {Opahle}},
  \bibinfo {author} {\bibfnamefont {H.~O.}\ \bibnamefont {Jeschke}},\ and\
  \bibinfo {author} {\bibfnamefont {R.}~\bibnamefont {Valent\'{\i}}},\ }\href
  {https://doi.org/10.1103/PhysRevB.81.094505} {\bibfield  {journal} {\bibinfo
  {journal} {Phys. Rev. B}\ }\textbf {\bibinfo {volume} {81}},\ \bibinfo
  {pages} {094505} (\bibinfo {year} {2010})}\BibitemShut {NoStop}%
\bibitem [{\citenamefont {Mazin}\ \emph {et~al.}(2008)\citenamefont {Mazin},
  \citenamefont {Singh}, \citenamefont {Johannes},\ and\ \citenamefont
  {Du}}]{Maz.08}%
  \BibitemOpen
  \bibfield  {author} {\bibinfo {author} {\bibfnamefont {I.~I.}\ \bibnamefont
  {Mazin}}, \bibinfo {author} {\bibfnamefont {D.~J.}\ \bibnamefont {Singh}},
  \bibinfo {author} {\bibfnamefont {M.~D.}\ \bibnamefont {Johannes}},\ and\
  \bibinfo {author} {\bibfnamefont {M.~H.}\ \bibnamefont {Du}},\ }\href
  {https://doi.org/10.1103/PhysRevLett.101.057003} {\bibfield  {journal}
  {\bibinfo  {journal} {Phys. Rev. Lett.}\ }\textbf {\bibinfo {volume} {101}},\
  \bibinfo {pages} {057003} (\bibinfo {year} {2008})}\BibitemShut {NoStop}%
\bibitem [{\citenamefont {Ning}\ \emph {et~al.}(2010)\citenamefont {Ning},
  \citenamefont {Ahilan}, \citenamefont {Imai}, \citenamefont {Sefat},
  \citenamefont {McGuire}, \citenamefont {Sales}, \citenamefont {Mandrus},
  \citenamefont {Cheng}, \citenamefont {Shen},\ and\ \citenamefont
  {Wen}}]{Nin.10}%
  \BibitemOpen
  \bibfield  {author} {\bibinfo {author} {\bibfnamefont {F.~L.}\ \bibnamefont
  {Ning}}, \bibinfo {author} {\bibfnamefont {K.}~\bibnamefont {Ahilan}},
  \bibinfo {author} {\bibfnamefont {T.}~\bibnamefont {Imai}}, \bibinfo {author}
  {\bibfnamefont {A.~S.}\ \bibnamefont {Sefat}}, \bibinfo {author}
  {\bibfnamefont {M.~A.}\ \bibnamefont {McGuire}}, \bibinfo {author}
  {\bibfnamefont {B.~C.}\ \bibnamefont {Sales}}, \bibinfo {author}
  {\bibfnamefont {D.}~\bibnamefont {Mandrus}}, \bibinfo {author} {\bibfnamefont
  {P.}~\bibnamefont {Cheng}}, \bibinfo {author} {\bibfnamefont
  {B.}~\bibnamefont {Shen}},\ and\ \bibinfo {author} {\bibfnamefont {H.-H.}\
  \bibnamefont {Wen}},\ }\href {https://doi.org/10.1103/PhysRevLett.104.037001}
  {\bibfield  {journal} {\bibinfo  {journal} {Phys. Rev. Lett.}\ }\textbf
  {\bibinfo {volume} {104}},\ \bibinfo {pages} {037001} (\bibinfo {year}
  {2010})}\BibitemShut {NoStop}%
\bibitem [{\citenamefont {Johnston}(2010)}]{Joh.10}%
  \BibitemOpen
  \bibfield  {author} {\bibinfo {author} {\bibfnamefont {D.~C.}\ \bibnamefont
  {Johnston}},\ }\href {https://doi.org/10.1080/00018732.2010.513480}
  {\bibfield  {journal} {\bibinfo  {journal} {Advances in Physics}\ }\textbf
  {\bibinfo {volume} {59}},\ \bibinfo {pages} {803} (\bibinfo {year} {2010})},\
  \Eprint {https://arxiv.org/abs/https://doi.org/10.1080/00018732.2010.513480}
  {https://doi.org/10.1080/00018732.2010.513480} \BibitemShut {NoStop}%
\bibitem [{\citenamefont {Si}\ and\ \citenamefont {Abrahams}(2008)}]{Si.08}%
  \BibitemOpen
  \bibfield  {author} {\bibinfo {author} {\bibfnamefont {Q.}~\bibnamefont
  {Si}}\ and\ \bibinfo {author} {\bibfnamefont {E.}~\bibnamefont {Abrahams}},\
  }\href {https://doi.org/10.1103/PhysRevLett.101.076401} {\bibfield  {journal}
  {\bibinfo  {journal} {Phys. Rev. Lett.}\ }\textbf {\bibinfo {volume} {101}},\
  \bibinfo {pages} {076401} (\bibinfo {year} {2008})}\BibitemShut {NoStop}%
\bibitem [{\citenamefont {Yildirim}(2008)}]{Yil.08}%
  \BibitemOpen
  \bibfield  {author} {\bibinfo {author} {\bibfnamefont {T.}~\bibnamefont
  {Yildirim}},\ }\href {https://doi.org/10.1103/PhysRevLett.101.057010}
  {\bibfield  {journal} {\bibinfo  {journal} {Phys. Rev. Lett.}\ }\textbf
  {\bibinfo {volume} {101}},\ \bibinfo {pages} {057010} (\bibinfo {year}
  {2008})}\BibitemShut {NoStop}%
\bibitem [{\citenamefont {Kou}\ \emph {et~al.}(2009)\citenamefont {Kou},
  \citenamefont {Li},\ and\ \citenamefont {Weng}}]{Kou.09}%
  \BibitemOpen
  \bibfield  {author} {\bibinfo {author} {\bibfnamefont {S.-P.}\ \bibnamefont
  {Kou}}, \bibinfo {author} {\bibfnamefont {T.}~\bibnamefont {Li}},\ and\
  \bibinfo {author} {\bibfnamefont {Z.-Y.}\ \bibnamefont {Weng}},\ }\href
  {https://doi.org/10.1209/0295-5075/88/17010} {\bibfield  {journal} {\bibinfo
  {journal} {{EPL} (Europhysics Letters)}\ }\textbf {\bibinfo {volume} {88}},\
  \bibinfo {pages} {17010} (\bibinfo {year} {2009})}\BibitemShut {NoStop}%
\bibitem [{\citenamefont {Li}\ \emph {et~al.}(2009)\citenamefont {Li},
  \citenamefont {de~la Cruz}, \citenamefont {Huang}, \citenamefont {Chen},
  \citenamefont {Xia}, \citenamefont {Luo}, \citenamefont {Wang},\ and\
  \citenamefont {Dai}}]{Li.09}%
  \BibitemOpen
  \bibfield  {author} {\bibinfo {author} {\bibfnamefont {S.}~\bibnamefont
  {Li}}, \bibinfo {author} {\bibfnamefont {C.}~\bibnamefont {de~la Cruz}},
  \bibinfo {author} {\bibfnamefont {Q.}~\bibnamefont {Huang}}, \bibinfo
  {author} {\bibfnamefont {G.~F.}\ \bibnamefont {Chen}}, \bibinfo {author}
  {\bibfnamefont {T.-L.}\ \bibnamefont {Xia}}, \bibinfo {author} {\bibfnamefont
  {J.~L.}\ \bibnamefont {Luo}}, \bibinfo {author} {\bibfnamefont {N.~L.}\
  \bibnamefont {Wang}},\ and\ \bibinfo {author} {\bibfnamefont
  {P.}~\bibnamefont {Dai}},\ }\href
  {https://doi.org/10.1103/PhysRevB.80.020504} {\bibfield  {journal} {\bibinfo
  {journal} {Phys. Rev. B}\ }\textbf {\bibinfo {volume} {80}},\ \bibinfo
  {pages} {020504} (\bibinfo {year} {2009})}\BibitemShut {NoStop}%
\bibitem [{\citenamefont {Akbari}\ \emph {et~al.}(2013)\citenamefont {Akbari},
  \citenamefont {Thalmeier},\ and\ \citenamefont {Eremin}}]{Akb.13}%
  \BibitemOpen
  \bibfield  {author} {\bibinfo {author} {\bibfnamefont {A.}~\bibnamefont
  {Akbari}}, \bibinfo {author} {\bibfnamefont {P.}~\bibnamefont {Thalmeier}},\
  and\ \bibinfo {author} {\bibfnamefont {I.}~\bibnamefont {Eremin}},\ }\href
  {https://doi.org/10.1088/1367-2630/15/3/033034} {\bibfield  {journal}
  {\bibinfo  {journal} {New Journal of Physics}\ }\textbf {\bibinfo {volume}
  {15}},\ \bibinfo {pages} {033034} (\bibinfo {year} {2013})}\BibitemShut
  {NoStop}%
\bibitem [{\citenamefont {Moriya}\ and\ \citenamefont {Ueda}(1974)}]{Mor.74}%
  \BibitemOpen
  \bibfield  {author} {\bibinfo {author} {\bibfnamefont {T.}~\bibnamefont
  {Moriya}}\ and\ \bibinfo {author} {\bibfnamefont {K.}~\bibnamefont {Ueda}},\
  }\href {https://doi.org/https://doi.org/10.1016/0038-1098(74)90733-9}
  {\bibfield  {journal} {\bibinfo  {journal} {Solid State Communications}\
  }\textbf {\bibinfo {volume} {15}},\ \bibinfo {pages} {169 } (\bibinfo {year}
  {1974})}\BibitemShut {NoStop}%
\bibitem [{\citenamefont {Song}\ \emph {et~al.}(2015)\citenamefont {Song},
  \citenamefont {Yamani}, \citenamefont {Cao}, \citenamefont {Li},
  \citenamefont {Zhang}, \citenamefont {Chen}, \citenamefont {Huang},
  \citenamefont {Wu}, \citenamefont {Tao}, \citenamefont {Zhu} \emph
  {et~al.}}]{Son.15}%
  \BibitemOpen
  \bibfield  {author} {\bibinfo {author} {\bibfnamefont {Y.}~\bibnamefont
  {Song}}, \bibinfo {author} {\bibfnamefont {Z.}~\bibnamefont {Yamani}},
  \bibinfo {author} {\bibfnamefont {C.}~\bibnamefont {Cao}}, \bibinfo {author}
  {\bibfnamefont {Y.}~\bibnamefont {Li}}, \bibinfo {author} {\bibfnamefont
  {C.}~\bibnamefont {Zhang}}, \bibinfo {author} {\bibfnamefont
  {J.}~\bibnamefont {Chen}}, \bibinfo {author} {\bibfnamefont {Q.}~\bibnamefont
  {Huang}}, \bibinfo {author} {\bibfnamefont {H.}~\bibnamefont {Wu}}, \bibinfo
  {author} {\bibfnamefont {J.}~\bibnamefont {Tao}}, \bibinfo {author}
  {\bibfnamefont {Y.}~\bibnamefont {Zhu}}, \emph {et~al.},\ }\href@noop {}
  {\bibfield  {journal} {\bibinfo  {journal} {arXiv preprint arXiv:1504.05116}\
  }\textbf {\bibinfo {volume} {7}} (\bibinfo {year} {2015})}\BibitemShut
  {NoStop}%
\bibitem [{\citenamefont {Ruderman}\ and\ \citenamefont
  {Kittel}(1954)}]{Rud.54}%
  \BibitemOpen
  \bibfield  {author} {\bibinfo {author} {\bibfnamefont {M.~A.}\ \bibnamefont
  {Ruderman}}\ and\ \bibinfo {author} {\bibfnamefont {C.}~\bibnamefont
  {Kittel}},\ }\href {https://doi.org/10.1103/PhysRev.96.99} {\bibfield
  {journal} {\bibinfo  {journal} {Phys. Rev.}\ }\textbf {\bibinfo {volume}
  {96}},\ \bibinfo {pages} {99} (\bibinfo {year} {1954})}\BibitemShut {NoStop}%
\bibitem [{\citenamefont {Boyce}\ and\ \citenamefont
  {Slichter}(1976)}]{Boy.76}%
  \BibitemOpen
  \bibfield  {author} {\bibinfo {author} {\bibfnamefont {J.~B.}\ \bibnamefont
  {Boyce}}\ and\ \bibinfo {author} {\bibfnamefont {C.~P.}\ \bibnamefont
  {Slichter}},\ }\href {https://doi.org/10.1103/PhysRevB.13.379} {\bibfield
  {journal} {\bibinfo  {journal} {Phys. Rev. B}\ }\textbf {\bibinfo {volume}
  {13}},\ \bibinfo {pages} {379} (\bibinfo {year} {1976})}\BibitemShut
  {NoStop}%
\bibitem [{\citenamefont {Jagannathan}\ \emph {et~al.}(1988)\citenamefont
  {Jagannathan}, \citenamefont {Abrahams},\ and\ \citenamefont
  {Stephen}}]{Jag.88}%
  \BibitemOpen
  \bibfield  {author} {\bibinfo {author} {\bibfnamefont {A.}~\bibnamefont
  {Jagannathan}}, \bibinfo {author} {\bibfnamefont {E.}~\bibnamefont
  {Abrahams}},\ and\ \bibinfo {author} {\bibfnamefont {M.~J.}\ \bibnamefont
  {Stephen}},\ }\href {https://doi.org/10.1103/PhysRevB.37.436} {\bibfield
  {journal} {\bibinfo  {journal} {Phys. Rev. B}\ }\textbf {\bibinfo {volume}
  {37}},\ \bibinfo {pages} {436} (\bibinfo {year} {1988})}\BibitemShut
  {NoStop}%
\bibitem [{\citenamefont {Kitagawa}\ \emph {et~al.}(2008)\citenamefont
  {Kitagawa}, \citenamefont {Katayama}, \citenamefont {Ohgushi}, \citenamefont
  {Yoshida},\ and\ \citenamefont {Takigawa}}]{Kit.08}%
  \BibitemOpen
  \bibfield  {author} {\bibinfo {author} {\bibfnamefont {K.}~\bibnamefont
  {Kitagawa}}, \bibinfo {author} {\bibfnamefont {N.}~\bibnamefont {Katayama}},
  \bibinfo {author} {\bibfnamefont {K.}~\bibnamefont {Ohgushi}}, \bibinfo
  {author} {\bibfnamefont {M.}~\bibnamefont {Yoshida}},\ and\ \bibinfo {author}
  {\bibfnamefont {M.}~\bibnamefont {Takigawa}},\ }\href
  {https://doi.org/10.1143/JPSJ.77.114709} {\bibfield  {journal} {\bibinfo
  {journal} {Journal of the Physical Society of Japan}\ }\textbf {\bibinfo
  {volume} {77}},\ \bibinfo {pages} {114709} (\bibinfo {year}
  {2008})}\BibitemShut {NoStop}%
\bibitem [{\citenamefont {Dioguardi}\ \emph {et~al.}(2010)\citenamefont
  {Dioguardi}, \citenamefont {apRoberts Warren}, \citenamefont {Shockley},
  \citenamefont {Bud'ko}, \citenamefont {Ni}, \citenamefont {Canfield},\ and\
  \citenamefont {Curro}}]{Dio.10}%
  \BibitemOpen
  \bibfield  {author} {\bibinfo {author} {\bibfnamefont {A.~P.}\ \bibnamefont
  {Dioguardi}}, \bibinfo {author} {\bibfnamefont {N.}~\bibnamefont {apRoberts
  Warren}}, \bibinfo {author} {\bibfnamefont {A.~C.}\ \bibnamefont {Shockley}},
  \bibinfo {author} {\bibfnamefont {S.~L.}\ \bibnamefont {Bud'ko}}, \bibinfo
  {author} {\bibfnamefont {N.}~\bibnamefont {Ni}}, \bibinfo {author}
  {\bibfnamefont {P.~C.}\ \bibnamefont {Canfield}},\ and\ \bibinfo {author}
  {\bibfnamefont {N.~J.}\ \bibnamefont {Curro}},\ }\href
  {https://doi.org/10.1103/PhysRevB.82.140411} {\bibfield  {journal} {\bibinfo
  {journal} {Phys. Rev. B}\ }\textbf {\bibinfo {volume} {82}},\ \bibinfo
  {pages} {140411} (\bibinfo {year} {2010})}\BibitemShut {NoStop}%
\bibitem [{\citenamefont {Dioguardi}\ \emph {et~al.}(2013)\citenamefont
  {Dioguardi}, \citenamefont {Crocker}, \citenamefont {Shockley}, \citenamefont
  {Lin}, \citenamefont {Shirer}, \citenamefont {Nisson}, \citenamefont
  {Lawson}, \citenamefont {Roberts-Warren}, \citenamefont {Canfield},
  \citenamefont {Bud'ko}, \citenamefont {Ran},\ and\ \citenamefont
  {Curro}}]{Dio.13}%
  \BibitemOpen
  \bibfield  {author} {\bibinfo {author} {\bibfnamefont {A.~P.}\ \bibnamefont
  {Dioguardi}}, \bibinfo {author} {\bibfnamefont {J.}~\bibnamefont {Crocker}},
  \bibinfo {author} {\bibfnamefont {A.~C.}\ \bibnamefont {Shockley}}, \bibinfo
  {author} {\bibfnamefont {C.~H.}\ \bibnamefont {Lin}}, \bibinfo {author}
  {\bibfnamefont {K.~R.}\ \bibnamefont {Shirer}}, \bibinfo {author}
  {\bibfnamefont {D.~M.}\ \bibnamefont {Nisson}}, \bibinfo {author}
  {\bibfnamefont {M.~M.}\ \bibnamefont {Lawson}}, \bibinfo {author}
  {\bibfnamefont {N.}~\bibnamefont {Roberts-Warren}}, \bibinfo {author}
  {\bibfnamefont {P.~C.}\ \bibnamefont {Canfield}}, \bibinfo {author}
  {\bibfnamefont {S.~L.}\ \bibnamefont {Bud'ko}}, \bibinfo {author}
  {\bibfnamefont {S.}~\bibnamefont {Ran}},\ and\ \bibinfo {author}
  {\bibfnamefont {N.~J.}\ \bibnamefont {Curro}},\ }\href
  {https://doi.org/10.1103/PhysRevLett.111.207201} {\bibfield  {journal}
  {\bibinfo  {journal} {Phys. Rev. Lett.}\ }\textbf {\bibinfo {volume} {111}},\
  \bibinfo {pages} {207201} (\bibinfo {year} {2013})}\BibitemShut {NoStop}%
\bibitem [{\citenamefont {Oh}\ \emph {et~al.}(2012)\citenamefont {Oh},
  \citenamefont {Mounce}, \citenamefont {Halperin}, \citenamefont {Zhang},
  \citenamefont {Dai}, \citenamefont {Reyes},\ and\ \citenamefont
  {Kuhns}}]{Oh.12}%
  \BibitemOpen
  \bibfield  {author} {\bibinfo {author} {\bibfnamefont {S.}~\bibnamefont
  {Oh}}, \bibinfo {author} {\bibfnamefont {A.~M.}\ \bibnamefont {Mounce}},
  \bibinfo {author} {\bibfnamefont {W.~P.}\ \bibnamefont {Halperin}}, \bibinfo
  {author} {\bibfnamefont {C.~L.}\ \bibnamefont {Zhang}}, \bibinfo {author}
  {\bibfnamefont {P.}~\bibnamefont {Dai}}, \bibinfo {author} {\bibfnamefont
  {A.~P.}\ \bibnamefont {Reyes}},\ and\ \bibinfo {author} {\bibfnamefont
  {P.~L.}\ \bibnamefont {Kuhns}},\ }\href
  {https://doi.org/10.1103/PhysRevB.85.174508} {\bibfield  {journal} {\bibinfo
  {journal} {Phys. Rev. B}\ }\textbf {\bibinfo {volume} {85}},\ \bibinfo
  {pages} {174508} (\bibinfo {year} {2012})}\BibitemShut {NoStop}%
\bibitem [{\citenamefont {Oh}\ \emph {et~al.}(2013)\citenamefont {Oh},
  \citenamefont {Mounce}, \citenamefont {Lee}, \citenamefont {Halperin},
  \citenamefont {Zhang}, \citenamefont {Carr},\ and\ \citenamefont
  {Dai}}]{Oh.13}%
  \BibitemOpen
  \bibfield  {author} {\bibinfo {author} {\bibfnamefont {S.}~\bibnamefont
  {Oh}}, \bibinfo {author} {\bibfnamefont {A.~M.}\ \bibnamefont {Mounce}},
  \bibinfo {author} {\bibfnamefont {J.~A.}\ \bibnamefont {Lee}}, \bibinfo
  {author} {\bibfnamefont {W.~P.}\ \bibnamefont {Halperin}}, \bibinfo {author}
  {\bibfnamefont {C.~L.}\ \bibnamefont {Zhang}}, \bibinfo {author}
  {\bibfnamefont {S.}~\bibnamefont {Carr}},\ and\ \bibinfo {author}
  {\bibfnamefont {P.}~\bibnamefont {Dai}},\ }\href
  {https://doi.org/10.1103/PhysRevB.87.174517} {\bibfield  {journal} {\bibinfo
  {journal} {Phys. Rev. B}\ }\textbf {\bibinfo {volume} {87}},\ \bibinfo
  {pages} {174517} (\bibinfo {year} {2013})}\BibitemShut {NoStop}%
\bibitem [{\citenamefont {Auler}\ \emph {et~al.}(1997)\citenamefont {Auler},
  \citenamefont {Horvati\ifmmode~\acute{c}\else \'{c}\fi{}}, \citenamefont
  {Gillet}, \citenamefont {Berthier}, \citenamefont {Berthier}, \citenamefont
  {S\'egransan},\ and\ \citenamefont {Henry}}]{Aul.97}%
  \BibitemOpen
  \bibfield  {author} {\bibinfo {author} {\bibfnamefont {T.}~\bibnamefont
  {Auler}}, \bibinfo {author} {\bibfnamefont {M.}~\bibnamefont
  {Horvati\ifmmode~\acute{c}\else \'{c}\fi{}}}, \bibinfo {author}
  {\bibfnamefont {J.~A.}\ \bibnamefont {Gillet}}, \bibinfo {author}
  {\bibfnamefont {C.}~\bibnamefont {Berthier}}, \bibinfo {author}
  {\bibfnamefont {Y.}~\bibnamefont {Berthier}}, \bibinfo {author}
  {\bibfnamefont {P.}~\bibnamefont {S\'egransan}},\ and\ \bibinfo {author}
  {\bibfnamefont {J.~Y.}\ \bibnamefont {Henry}},\ }\href
  {https://doi.org/10.1103/PhysRevB.56.11294} {\bibfield  {journal} {\bibinfo
  {journal} {Phys. Rev. B}\ }\textbf {\bibinfo {volume} {56}},\ \bibinfo
  {pages} {11294} (\bibinfo {year} {1997})}\BibitemShut {NoStop}%
\bibitem [{\citenamefont {Tan}\ \emph {et~al.}(2017)\citenamefont {Tan},
  \citenamefont {Song}, \citenamefont {Zhang}, \citenamefont {Lin},
  \citenamefont {Xu}, \citenamefont {Tian}, \citenamefont {Chi}, \citenamefont
  {Graves-Brook}, \citenamefont {Li},\ and\ \citenamefont {Dai}}]{Tan.17}%
  \BibitemOpen
  \bibfield  {author} {\bibinfo {author} {\bibfnamefont {G.}~\bibnamefont
  {Tan}}, \bibinfo {author} {\bibfnamefont {Y.}~\bibnamefont {Song}}, \bibinfo
  {author} {\bibfnamefont {R.}~\bibnamefont {Zhang}}, \bibinfo {author}
  {\bibfnamefont {L.}~\bibnamefont {Lin}}, \bibinfo {author} {\bibfnamefont
  {Z.}~\bibnamefont {Xu}}, \bibinfo {author} {\bibfnamefont {L.}~\bibnamefont
  {Tian}}, \bibinfo {author} {\bibfnamefont {S.}~\bibnamefont {Chi}}, \bibinfo
  {author} {\bibfnamefont {M.~K.}\ \bibnamefont {Graves-Brook}}, \bibinfo
  {author} {\bibfnamefont {S.}~\bibnamefont {Li}},\ and\ \bibinfo {author}
  {\bibfnamefont {P.}~\bibnamefont {Dai}},\ }\href
  {https://doi.org/10.1103/PhysRevB.95.054501} {\bibfield  {journal} {\bibinfo
  {journal} {Phys. Rev. B}\ }\textbf {\bibinfo {volume} {95}},\ \bibinfo
  {pages} {054501} (\bibinfo {year} {2017})}\BibitemShut {NoStop}%
\bibitem [{\citenamefont {Charnukha}\ \emph {et~al.}(2017)\citenamefont
  {Charnukha}, \citenamefont {Yin}, \citenamefont {Song}, \citenamefont {Cao},
  \citenamefont {Dai}, \citenamefont {Haule}, \citenamefont {Kotliar},\ and\
  \citenamefont {Basov}}]{Cha.17}%
  \BibitemOpen
  \bibfield  {author} {\bibinfo {author} {\bibfnamefont {A.}~\bibnamefont
  {Charnukha}}, \bibinfo {author} {\bibfnamefont {Z.~P.}\ \bibnamefont {Yin}},
  \bibinfo {author} {\bibfnamefont {Y.}~\bibnamefont {Song}}, \bibinfo {author}
  {\bibfnamefont {C.~D.}\ \bibnamefont {Cao}}, \bibinfo {author} {\bibfnamefont
  {P.}~\bibnamefont {Dai}}, \bibinfo {author} {\bibfnamefont {K.}~\bibnamefont
  {Haule}}, \bibinfo {author} {\bibfnamefont {G.}~\bibnamefont {Kotliar}},\
  and\ \bibinfo {author} {\bibfnamefont {D.~N.}\ \bibnamefont {Basov}},\ }\href
  {https://doi.org/10.1103/PhysRevB.96.195121} {\bibfield  {journal} {\bibinfo
  {journal} {Phys. Rev. B}\ }\textbf {\bibinfo {volume} {96}},\ \bibinfo
  {pages} {195121} (\bibinfo {year} {2017})}\BibitemShut {NoStop}%
\bibitem [{\citenamefont {Curro}\ \emph {et~al.}(2000)\citenamefont {Curro},
  \citenamefont {Hammel}, \citenamefont {Suh}, \citenamefont {H\"ucker},
  \citenamefont {B\"uchner}, \citenamefont {Ammerahl},\ and\ \citenamefont
  {Revcolevschi}}]{Cur.00}%
  \BibitemOpen
  \bibfield  {author} {\bibinfo {author} {\bibfnamefont {N.~J.}\ \bibnamefont
  {Curro}}, \bibinfo {author} {\bibfnamefont {P.~C.}\ \bibnamefont {Hammel}},
  \bibinfo {author} {\bibfnamefont {B.~J.}\ \bibnamefont {Suh}}, \bibinfo
  {author} {\bibfnamefont {M.}~\bibnamefont {H\"ucker}}, \bibinfo {author}
  {\bibfnamefont {B.}~\bibnamefont {B\"uchner}}, \bibinfo {author}
  {\bibfnamefont {U.}~\bibnamefont {Ammerahl}},\ and\ \bibinfo {author}
  {\bibfnamefont {A.}~\bibnamefont {Revcolevschi}},\ }\href
  {https://doi.org/10.1103/PhysRevLett.85.642} {\bibfield  {journal} {\bibinfo
  {journal} {Phys. Rev. Lett.}\ }\textbf {\bibinfo {volume} {85}},\ \bibinfo
  {pages} {642} (\bibinfo {year} {2000})}\BibitemShut {NoStop}%
\bibitem [{\citenamefont {Kemper}\ \emph {et~al.}(2009)\citenamefont {Kemper},
  \citenamefont {Cao}, \citenamefont {Hirschfeld},\ and\ \citenamefont
  {Cheng}}]{Kem.09}%
  \BibitemOpen
  \bibfield  {author} {\bibinfo {author} {\bibfnamefont {A.~F.}\ \bibnamefont
  {Kemper}}, \bibinfo {author} {\bibfnamefont {C.}~\bibnamefont {Cao}},
  \bibinfo {author} {\bibfnamefont {P.~J.}\ \bibnamefont {Hirschfeld}},\ and\
  \bibinfo {author} {\bibfnamefont {H.-P.}\ \bibnamefont {Cheng}},\ }\href
  {https://doi.org/10.1103/PhysRevB.80.104511} {\bibfield  {journal} {\bibinfo
  {journal} {Phys. Rev. B}\ }\textbf {\bibinfo {volume} {80}},\ \bibinfo
  {pages} {104511} (\bibinfo {year} {2009})}\BibitemShut {NoStop}%
\bibitem [{\citenamefont {Wadati}\ \emph {et~al.}(2010)\citenamefont {Wadati},
  \citenamefont {Elfimov},\ and\ \citenamefont {Sawatzky}}]{Wad.10}%
  \BibitemOpen
  \bibfield  {author} {\bibinfo {author} {\bibfnamefont {H.}~\bibnamefont
  {Wadati}}, \bibinfo {author} {\bibfnamefont {I.}~\bibnamefont {Elfimov}},\
  and\ \bibinfo {author} {\bibfnamefont {G.~A.}\ \bibnamefont {Sawatzky}},\
  }\href {https://doi.org/10.1103/PhysRevLett.105.157004} {\bibfield  {journal}
  {\bibinfo  {journal} {Phys. Rev. Lett.}\ }\textbf {\bibinfo {volume} {105}},\
  \bibinfo {pages} {157004} (\bibinfo {year} {2010})}\BibitemShut {NoStop}%
\bibitem [{\citenamefont {Zhou}\ \emph {et~al.}(2012)\citenamefont {Zhou},
  \citenamefont {Cai}, \citenamefont {Wang}, \citenamefont {Ruan},
  \citenamefont {Ye}, \citenamefont {Chen}, \citenamefont {You}, \citenamefont
  {Weng},\ and\ \citenamefont {Wang}}]{Zho.12}%
  \BibitemOpen
  \bibfield  {author} {\bibinfo {author} {\bibfnamefont {X.}~\bibnamefont
  {Zhou}}, \bibinfo {author} {\bibfnamefont {P.}~\bibnamefont {Cai}}, \bibinfo
  {author} {\bibfnamefont {A.}~\bibnamefont {Wang}}, \bibinfo {author}
  {\bibfnamefont {W.}~\bibnamefont {Ruan}}, \bibinfo {author} {\bibfnamefont
  {C.}~\bibnamefont {Ye}}, \bibinfo {author} {\bibfnamefont {X.}~\bibnamefont
  {Chen}}, \bibinfo {author} {\bibfnamefont {Y.}~\bibnamefont {You}}, \bibinfo
  {author} {\bibfnamefont {Z.-Y.}\ \bibnamefont {Weng}},\ and\ \bibinfo
  {author} {\bibfnamefont {Y.}~\bibnamefont {Wang}},\ }\href
  {https://doi.org/10.1103/PhysRevLett.109.037002} {\bibfield  {journal}
  {\bibinfo  {journal} {Phys. Rev. Lett.}\ }\textbf {\bibinfo {volume} {109}},\
  \bibinfo {pages} {037002} (\bibinfo {year} {2012})}\BibitemShut {NoStop}%
\bibitem [{\citenamefont {Tan}\ \emph {et~al.}(2016)\citenamefont {Tan},
  \citenamefont {Song}, \citenamefont {Zhang}, \citenamefont {Lin},
  \citenamefont {Xu}, \citenamefont {Hou}, \citenamefont {Tian}, \citenamefont
  {Cao}, \citenamefont {Li}, \citenamefont {Feng},\ and\ \citenamefont
  {Dai}}]{Tan.16}%
  \BibitemOpen
  \bibfield  {author} {\bibinfo {author} {\bibfnamefont {G.}~\bibnamefont
  {Tan}}, \bibinfo {author} {\bibfnamefont {Y.}~\bibnamefont {Song}}, \bibinfo
  {author} {\bibfnamefont {C.}~\bibnamefont {Zhang}}, \bibinfo {author}
  {\bibfnamefont {L.}~\bibnamefont {Lin}}, \bibinfo {author} {\bibfnamefont
  {Z.}~\bibnamefont {Xu}}, \bibinfo {author} {\bibfnamefont {T.}~\bibnamefont
  {Hou}}, \bibinfo {author} {\bibfnamefont {W.}~\bibnamefont {Tian}}, \bibinfo
  {author} {\bibfnamefont {H.}~\bibnamefont {Cao}}, \bibinfo {author}
  {\bibfnamefont {S.}~\bibnamefont {Li}}, \bibinfo {author} {\bibfnamefont
  {S.}~\bibnamefont {Feng}},\ and\ \bibinfo {author} {\bibfnamefont
  {P.}~\bibnamefont {Dai}},\ }\href
  {https://doi.org/10.1103/PhysRevB.94.014509} {\bibfield  {journal} {\bibinfo
  {journal} {Phys. Rev. B}\ }\textbf {\bibinfo {volume} {94}},\ \bibinfo
  {pages} {014509} (\bibinfo {year} {2016})}\BibitemShut {NoStop}%
\bibitem [{\citenamefont {Ideta}\ \emph {et~al.}(2013)\citenamefont {Ideta},
  \citenamefont {Yoshida}, \citenamefont {Nishi}, \citenamefont {Fujimori},
  \citenamefont {Kotani}, \citenamefont {Ono}, \citenamefont {Nakashima},
  \citenamefont {Yamaichi}, \citenamefont {Sasagawa}, \citenamefont {Nakajima},
  \citenamefont {Kihou}, \citenamefont {Tomioka}, \citenamefont {Lee},
  \citenamefont {Iyo}, \citenamefont {Eisaki}, \citenamefont {Ito},
  \citenamefont {Uchida},\ and\ \citenamefont {Arita}}]{Ide.13}%
  \BibitemOpen
  \bibfield  {author} {\bibinfo {author} {\bibfnamefont {S.}~\bibnamefont
  {Ideta}}, \bibinfo {author} {\bibfnamefont {T.}~\bibnamefont {Yoshida}},
  \bibinfo {author} {\bibfnamefont {I.}~\bibnamefont {Nishi}}, \bibinfo
  {author} {\bibfnamefont {A.}~\bibnamefont {Fujimori}}, \bibinfo {author}
  {\bibfnamefont {Y.}~\bibnamefont {Kotani}}, \bibinfo {author} {\bibfnamefont
  {K.}~\bibnamefont {Ono}}, \bibinfo {author} {\bibfnamefont {Y.}~\bibnamefont
  {Nakashima}}, \bibinfo {author} {\bibfnamefont {S.}~\bibnamefont {Yamaichi}},
  \bibinfo {author} {\bibfnamefont {T.}~\bibnamefont {Sasagawa}}, \bibinfo
  {author} {\bibfnamefont {M.}~\bibnamefont {Nakajima}}, \bibinfo {author}
  {\bibfnamefont {K.}~\bibnamefont {Kihou}}, \bibinfo {author} {\bibfnamefont
  {Y.}~\bibnamefont {Tomioka}}, \bibinfo {author} {\bibfnamefont {C.~H.}\
  \bibnamefont {Lee}}, \bibinfo {author} {\bibfnamefont {A.}~\bibnamefont
  {Iyo}}, \bibinfo {author} {\bibfnamefont {H.}~\bibnamefont {Eisaki}},
  \bibinfo {author} {\bibfnamefont {T.}~\bibnamefont {Ito}}, \bibinfo {author}
  {\bibfnamefont {S.}~\bibnamefont {Uchida}},\ and\ \bibinfo {author}
  {\bibfnamefont {R.}~\bibnamefont {Arita}},\ }\href
  {https://doi.org/10.1103/PhysRevLett.110.107007} {\bibfield  {journal}
  {\bibinfo  {journal} {Phys. Rev. Lett.}\ }\textbf {\bibinfo {volume} {110}},\
  \bibinfo {pages} {107007} (\bibinfo {year} {2013})}\BibitemShut {NoStop}%
\bibitem [{\citenamefont {Wang}\ \emph {et~al.}(2018)\citenamefont {Wang},
  \citenamefont {Song}, \citenamefont {Cao}, \citenamefont {Tseng},
  \citenamefont {Keller}, \citenamefont {Li}, \citenamefont {Harriger},
  \citenamefont {Tian}, \citenamefont {Chi}, \citenamefont {Yu} \emph
  {et~al.}}]{Wan.18}%
  \BibitemOpen
  \bibfield  {author} {\bibinfo {author} {\bibfnamefont {W.}~\bibnamefont
  {Wang}}, \bibinfo {author} {\bibfnamefont {Y.}~\bibnamefont {Song}}, \bibinfo
  {author} {\bibfnamefont {C.}~\bibnamefont {Cao}}, \bibinfo {author}
  {\bibfnamefont {K.-F.}\ \bibnamefont {Tseng}}, \bibinfo {author}
  {\bibfnamefont {T.}~\bibnamefont {Keller}}, \bibinfo {author} {\bibfnamefont
  {Y.}~\bibnamefont {Li}}, \bibinfo {author} {\bibfnamefont {L.~W.}\
  \bibnamefont {Harriger}}, \bibinfo {author} {\bibfnamefont {W.}~\bibnamefont
  {Tian}}, \bibinfo {author} {\bibfnamefont {S.}~\bibnamefont {Chi}}, \bibinfo
  {author} {\bibfnamefont {R.}~\bibnamefont {Yu}}, \emph {et~al.},\ }\href@noop
  {} {\bibfield  {journal} {\bibinfo  {journal} {Nature communications}\
  }\textbf {\bibinfo {volume} {9}},\ \bibinfo {pages} {1} (\bibinfo {year}
  {2018})}\BibitemShut {NoStop}%
\bibitem [{\citenamefont {Galsin}(2019)}]{Gal.19}%
  \BibitemOpen
  \bibfield  {author} {\bibinfo {author} {\bibfnamefont {J.}~\bibnamefont
  {Galsin}},\ }\href {https://books.google.com/books?id=mNGJDwAAQBAJ} {\emph
  {\bibinfo {title} {{Solid State Physics: An Introduction to Theory}}}}\
  (\bibinfo  {publisher} {Elsevier Science},\ \bibinfo {year}
  {2019})\BibitemShut {NoStop}%
\bibitem [{\citenamefont {Yin}\ \emph {et~al.}(2010)\citenamefont {Yin},
  \citenamefont {Lee},\ and\ \citenamefont {Ku}}]{Yin.10}%
  \BibitemOpen
  \bibfield  {author} {\bibinfo {author} {\bibfnamefont {W.-G.}\ \bibnamefont
  {Yin}}, \bibinfo {author} {\bibfnamefont {C.-C.}\ \bibnamefont {Lee}},\ and\
  \bibinfo {author} {\bibfnamefont {W.}~\bibnamefont {Ku}},\ }\href
  {https://doi.org/10.1103/PhysRevLett.105.107004} {\bibfield  {journal}
  {\bibinfo  {journal} {Phys. Rev. Lett.}\ }\textbf {\bibinfo {volume} {105}},\
  \bibinfo {pages} {107004} (\bibinfo {year} {2010})}\BibitemShut {NoStop}%
\bibitem [{\citenamefont {Wu}\ \emph {et~al.}(2008)\citenamefont {Wu},
  \citenamefont {Phillips},\ and\ \citenamefont {Castro~Neto}}]{Wu.08}%
  \BibitemOpen
  \bibfield  {author} {\bibinfo {author} {\bibfnamefont {J.}~\bibnamefont
  {Wu}}, \bibinfo {author} {\bibfnamefont {P.}~\bibnamefont {Phillips}},\ and\
  \bibinfo {author} {\bibfnamefont {A.~H.}\ \bibnamefont {Castro~Neto}},\
  }\href {https://doi.org/10.1103/PhysRevLett.101.126401} {\bibfield  {journal}
  {\bibinfo  {journal} {Phys. Rev. Lett.}\ }\textbf {\bibinfo {volume} {101}},\
  \bibinfo {pages} {126401} (\bibinfo {year} {2008})}\BibitemShut {NoStop}%
\bibitem [{\citenamefont {de’Medici}\ \emph {et~al.}(2009)\citenamefont
  {de’Medici}, \citenamefont {Hassan},\ and\ \citenamefont
  {Capone}}]{Med.09}%
  \BibitemOpen
  \bibfield  {author} {\bibinfo {author} {\bibfnamefont {L.}~\bibnamefont
  {de’Medici}}, \bibinfo {author} {\bibfnamefont {S.~R.}\ \bibnamefont
  {Hassan}},\ and\ \bibinfo {author} {\bibfnamefont {M.}~\bibnamefont
  {Capone}},\ }\href@noop {} {\bibfield  {journal} {\bibinfo  {journal}
  {Journal of superconductivity and novel magnetism}\ }\textbf {\bibinfo
  {volume} {22}},\ \bibinfo {pages} {535} (\bibinfo {year} {2009})}\BibitemShut
  {NoStop}%
\bibitem [{\citenamefont {Mitrovi\ifmmode~\acute{c}\else \'{c}\fi{}}\ \emph
  {et~al.}(2008)\citenamefont {Mitrovi\ifmmode~\acute{c}\else \'{c}\fi{}},
  \citenamefont {Julien}, \citenamefont {de~Vaulx}, \citenamefont
  {Horvati\ifmmode~\acute{c}\else \'{c}\fi{}}, \citenamefont {Berthier},
  \citenamefont {Suzuki},\ and\ \citenamefont {Yamada}}]{Mit.08}%
  \BibitemOpen
  \bibfield  {author} {\bibinfo {author} {\bibfnamefont {V.~F.}\ \bibnamefont
  {Mitrovi\ifmmode~\acute{c}\else \'{c}\fi{}}}, \bibinfo {author}
  {\bibfnamefont {M.-H.}\ \bibnamefont {Julien}}, \bibinfo {author}
  {\bibfnamefont {C.}~\bibnamefont {de~Vaulx}}, \bibinfo {author}
  {\bibfnamefont {M.}~\bibnamefont {Horvati\ifmmode~\acute{c}\else
  \'{c}\fi{}}}, \bibinfo {author} {\bibfnamefont {C.}~\bibnamefont {Berthier}},
  \bibinfo {author} {\bibfnamefont {T.}~\bibnamefont {Suzuki}},\ and\ \bibinfo
  {author} {\bibfnamefont {K.}~\bibnamefont {Yamada}},\ }\href
  {https://doi.org/10.1103/PhysRevB.78.014504} {\bibfield  {journal} {\bibinfo
  {journal} {Phys. Rev. B}\ }\textbf {\bibinfo {volume} {78}},\ \bibinfo
  {pages} {014504} (\bibinfo {year} {2008})}\BibitemShut {NoStop}%
\bibitem [{\citenamefont {Johnston}(2006)}]{Joh.06}%
  \BibitemOpen
  \bibfield  {author} {\bibinfo {author} {\bibfnamefont {D.~C.}\ \bibnamefont
  {Johnston}},\ }\href {https://doi.org/10.1103/PhysRevB.74.184430} {\bibfield
  {journal} {\bibinfo  {journal} {Phys. Rev. B}\ }\textbf {\bibinfo {volume}
  {74}},\ \bibinfo {pages} {184430} (\bibinfo {year} {2006})}\BibitemShut
  {NoStop}%
\end{thebibliography}%
\end{document}